\documentclass[11pt]{article}

\usepackage{amssymb}
\usepackage{tikz}
\usepackage{amsmath}

\usepackage{graphicx}

\usepackage{mathtools}

\newtagform{brackets}{(}{)}
\usetagform{brackets}

\usepackage{dsfont}

\usepackage{ragged2e}
\usepackage{enumitem}
\usepackage{etoolbox}
\usepackage{booktabs}

\usepackage{color}

\usepackage{amsfonts}

\usepackage{xfrac}
\usepackage{hyperref}
\usepackage{booktabs}
\usepackage[makeroom]{cancel}

\def\p{\partial}

\newcommand{\be}{\begin{eqnarray}}
\newcommand{\ee}{\end{eqnarray}}

\def\p{\partial}

\begin{document}

\title{6D dual superconformal algebra}

\author{{\sc Klaus~Bering}$^{a}$ and {\sc Michal~Pazderka}$^{b}$ \\~\\
Institute for Theoretical Physics \& Astrophysics\\
Masaryk University\\Kotl\'a\v{r}sk\'a 2\\CZ--611 37 Brno\\Czech Republic}


\maketitle

\begin{abstract}
We construct and study the 6D dual superconformal algebra. Our construction is inspired by the dual superconformal symmetry of massless 4D $\mathcal{N}=4$ SYM and extends the previous construction of the enhanced dual conformal algebra for 6D $\mathcal{N}=(1,1)$ SYM to the  full 6D dual superconformal algebra for chiral theories. We formulate constraints in 6D spinor helicity formalism and find all generators of the 6D dual superconformal algebra. Next we check that they agree with the dual superconformal generators of known 3D and 4D theories. We show that it is possible to significantly simplify the form of generators and compactly write the dual superconformal algebra using superindices. Finally, we work out some examples of algebra invariants.
\end{abstract}

\vfill
\begin{quote}
\hrule width 5.cm \vskip 2.mm \noindent 
$^{a}${\small E--mail:~{\tt bering@physics.muni.cz}} \hspace{10mm}
$^{b}${\small E--mail:~{\tt michpa@mail.muni.cz}} \\
\end{quote}
\newpage

\section{Introduction}
Scattering amplitudes are one of the most important and most interesting outputs of quantum field theories. The standard procedure how to calculate scattering amplitudes is via Feynman diagrams, which is a perturbative approach. However, the complexity of calculations rapidly grows with the number of external legs and with the number of loops. Despite the complexity of calculations a huge number of intermediate terms often cancel and the final scattering amplitude can be extremely simple, e.g.\ tree-level gluon amplitudes, cf.\ Parke \& Taylor \cite{Parke_Taylor}. This might suggest some deeper structure. 

There has been great progress in exploring scattering amplitudes in recent years, see for example \cite{BCFW05}, \cite{Witten04}, \cite{Trnka12} and many more.  One of the most interesting theories is the massless $D=4$ $\mathcal{N}=4$ SYM, where the underlying structure is the Yangian symmetry. In detail this theory is conformally invariant not only in the bosonic sector but it enjoys full superconformal symmetry. Apart from the superconformal symmetry there is also a hidden symmetry called dual superconformal symmetry \cite{Drummond08}. Taking commutators of generators of these algebras give rise to infinitely many new generators forming a Yangian algebra and therefore it leads to Yangian symmetry of $\mathcal{N}=4$ SYM \cite{Plefka09}.

It is an interesting question how much of this symmetry survives for massive amplitudes i.e.\ scattering amplitudes on the Coulomb branch.  Although this theory is no longer conformally invariant it turns out to be still dual conformally invariant \cite{Dennen_10}, \cite{Plefka14}. The questions of potential dual superconformal invariance and Yangian-like invariance are still open which motivates us to construct the 6D dual superconformal algebra. Apart from 4D  massive amplitudes, the implications of the 6D dual superconformal algebra for scattering amplitudes might be interesting by themselves. For example the 6D dual conformal symmetry fixes the 1-loop 4-point integrand \cite{Lipstein16}.  

In this paper we want to build a general framework for the dual superconformal algebra in various dimensions which can be immediately applied to massive scattering amplitudes in 4D via dimensional reduction. The paper is organized as follows in terms of chapters: II.\ We establish the necessary spinor helicity formalism in 6D. III.\ We briefly review the dual superconformal symmetry of massless $D=4$ $\mathcal{N}=4$ SYM. IV.\ We construct the dual superconformal algebra in 6D in analogy with the 4D case. V.\ A significant simplification of calculations is achieved by introducing superindices, and as a result, expressions for all generators become very compact. VI.\ Applying dimensional reduction to $D=4$ and $D=3$ we show that our results reduce to known generators of dual superconformal algebra (DSCA) in appropriate dimensions. VII.\ We briefly discuss how an analogue of the 4D central charge sits in our formalism. VIII.\ We work out some examples of algebra invariants.

\section{Spinor helicity formalism in 6D}\label{6Dspinors}
Spinor helicity formalism turns out to be a very powerful tool for describing scattering amplitudes effectively \cite{Dixon09}, \cite{Cheung09}. The main advantage of using spinor helicity formalism is to use spinor variables adopted for massless particles. Therefore we do not have to impose extra constraints (e.g.\ on polarization vectors) and as a consequence scattering amplitudes take a much simpler form. It can be done by  solving  the equation for massless particles 
\be
\label{mass_less}
p^2~=~0~.
\ee
\indent Let us first discuss the Lorentz group in 6D which is $SO^+(1,5)$ and the corresponding spin group which is $Spin(1,5)$. It is more convenient to use the complexification of $Spin(1,5)$ which is $Spin(6,\mathbb{C})\cong SU(4)_{\mathbb{C}}\cong SL(4,\mathbb{C})$ because the antisymmetric representation of $SU(4)_{\mathbb{C}}$ is six-dimensional. In other words vectors are antisymmetric in spinor indices. 
\\
\indent  This can be viewed in the following way. First let's replace the 6D spacetime index $\mu$ with spinor indices $AB$ (with the help of the 6D Pauli matrices)
\be
\left(\Sigma^{\mu}\right)^{AB}p_{i\mu}=p^{AB}_i\quad\longleftarrow \qquad p_{i\mu}\qquad  \longrightarrow \quad\tilde{p}_{iAB}=\left(\tilde{\Sigma}^{\mu}\right)_{AB}p_{i\mu}~,
\ee
where $\tilde{p}_{AB}$ and $p^{AB}$ are antisymmetric in $A$ and $B$, cf. Table \ref{table1}.
These two representations of momentum $p^{AB}$ and $\tilde{p}_{AB}$ are not independent because the $(\Sigma^{\mu})^{AB}$ and $(\tilde{\Sigma}^{\mu})_{AB}$ matrices are not independent but related via the $SU(4)$ metric (= the totally antisymmetric Levi-Civita symbol with four indices)
\be
(\Sigma^{\mu})^{AB}=\frac{1}{2}\epsilon^{ABCD}(\tilde{\Sigma}^{\mu})_{CD} \quad \Rightarrow \quad p^{AB}_i=\frac{1}{2}\epsilon^{ABCD}\tilde{p}_{iCD}~.
\ee
We can then write the null momentum 
\be
p^{AB}_i=\lambda^{Aa}_i\lambda^{B}_{ia}~, \qquad \tilde{p}_{iAB}=\tilde{\lambda}_{iA\dot{a}}\tilde{\lambda}_{iB}^{\dot{a}}~,
\ee
using spinors $\lambda^{Aa}_i$ or $\tilde{\lambda}_{iA\dot{a}}$, where $A$ is  a $SU(4)$ spinor index and $a$ and $\dot{a}$ are little group indices corresponding to the little group $SO(4) \cong [SU(2)\times SU(2)]/\mathbb{Z}_2$.  

\indent We can think about spinors also as solutions to the massless Dirac equations
\be
\label{Dirac}
\tilde{p}_{AB}\lambda^{Aa}=0~, \qquad p^{AB}\tilde{\lambda}_{A\dot{a}}=0~.
\ee
The momentum as a complex antisymmetric matrix $p^{AB}$ has rank two for massless particles.  (This is due to $\det (p^{AB})=(p^2)^2=0$ for massless particles and the fact that the rank of an antisymmetric complex matrix is an even number \cite{Horn}.) Therefore we have two independent solutions to each of Dirac equations $\tilde{p}_{AB}\lambda^{Aa}=0$ and $p^{AB}\tilde{\lambda}_{A\dot{a}}=0$ labeled by the little group indices. Sometimes it is useful to have an explicit expression of spinors in terms of momenta which can be achieved via solving Dirac equations \eqref{Dirac}. One possible representation is \cite{Plefka14}
\be
\label{Lambda_6D}
\lambda^{Aa}=\left(
\begin{array}{cc}
0& \sqrt{p_0+p_3}\\
\frac{-p_5+ip_4}{\sqrt{p_0+p_3}}& \frac{p_1+ip_2}{\sqrt{p_0+p_3}}\\
\frac{-p_1+ip_2}{\sqrt{p_0+p_3}}& \frac{-p_5-ip_4}{\sqrt{p_0+p_3}}\\
\sqrt{p_0+p_3}&0
\end{array}
\right)~, \quad \tilde{\lambda}_{A\dot{a}}=\left(
\begin{array}{cc}
0& \sqrt{p_0-p_3}\\
\frac{p_5+ip_4}{\sqrt{p_0-p_3}}& \frac{-p_1+ip_2}{\sqrt{p_0-p_3}}\\
\frac{p_1+ip_2}{\sqrt{p_0-p_3}}& \frac{p_5-ip_4}{\sqrt{p_0-p_3}}\\
\sqrt{p_0-p_3}&0
\end{array}
\right)~.
\ee 
For more details, see for example \cite{Cheung09}. For spinor helicity formalism in 4D,  see for example \cite{Dixon09}. \\
\indent Table \ref{table1} contains a list of letters and symbols that
we will use in the text for different types of indices.
\begin{table}[h!]
\begin{center}
\caption{Index conventions}
\label{table1}
\begin{tabular}{l|l|l|l}
$\textbf{Index}$&$\textbf{6D}$&$\textbf{4D}$&$\textbf{3D}$\\
\hline	
Lorentz&$\mu,\nu$&$\mu,\nu$&$\mu,\nu$\\
Spinor&$A, B, C, D, E, F, G, H$&$\alpha, \dot{\alpha}, \beta, \dot{\beta},\gamma, \dot{\gamma}$& $\alpha, \beta, \gamma$\\
&&$\delta, \dot{\delta}, \epsilon, \dot{\epsilon}$&\\
Little group&$a,\dot{a},b,\dot{b}, c ,\dot{c}$&-&-\\
Particle&$i,j,k$&$i,j,k$&$i,j,k$\\

$R$-symmetry&$I,J,K,L,M,N$&$I,J,K,L,M$&$I,J,K,L,M$\\
\end{tabular}
\end{center}
\end{table}

\section{Dual superconformal symmetry of massless $D=4$ $\mathcal{N}=4$  SYM}

\indent Scattering amplitudes of $\mathcal{N}=4$ SYM take a very simple form. An example of such superamplitudes is the well known tree-level  $MHV$ scattering amplitude 
\be
\label{MHV-amp}
\mathcal{A}_n^{MHV}=\frac{\delta^4\left(p^{\alpha\dot{\alpha}}\right)\delta^{8}\left(q^{I\alpha}\right)}{\left<12\right>\ldots \left<n1\right>}~.
\ee
Before reviewing  the dual superconformal algebra it is useful to recall that tree-level scattering amplitudes enjoy superconformal symmetry generated by the following set of generators
\be
\label{g_4D}
g_{4D}=\{p^{\alpha\dot{\alpha}}, m_{\alpha\beta}, \bar{m}_{\dot{\alpha}\dot{\beta}}, d, k_{\alpha\dot{\alpha}}, q^{I\alpha}, \bar{q}^{\dot{\alpha}}_I, s_{I\alpha}, \bar{s}^I_{\dot{\alpha}}, r^I{}_J, c\}~,
\ee
which form a $psu(2,2|4)$ algebra. In this text the generators of the superconformal algebra are labeled by small letters and generators of dual algebra are labeled by capital letters. The explicit form of superconformal generators can be found for example in \cite{Plefka09}.\\
\indent Moreover scattering amplitudes enjoy a hidden dual superconformal symmetry. This symmetry was discovered for the first time in \cite{Drummond08} and puts a nontrivial constraints on scattering amplitudes. It is convenient to express generators of the dual superconformal symmetry  (DSCS) with help of  the dual coordinates $x$ and $\theta$ satisfying  the following constraints\footnote{\label{footnote4} The symbol $\approx$ means weakly equal, i.e.\ the left and right hand sides of this symbol are equal up to constraints.}
\begin{equation}
\label{C_4d}
x_i^{\alpha\dot{\alpha}}-x_{i+1}^{\alpha\dot{\alpha}}-\lambda_i^{\alpha}\tilde{\lambda}_i^{\dot{\alpha}}\approx 0 ~,
\end{equation}
\begin{equation*}
\theta_i^{I\alpha}-\theta_{i+1}^{I\alpha}-\eta_i^I\lambda_i^{\alpha}\approx0~,
\end{equation*}
 i.e.  eqs.\ \eqref{C_4d} define a subspace in the so-called \textit{full space} $\{\lambda_i^{\alpha}, \tilde{\lambda}_i^{\dot{\alpha}}, \eta_i^{I}, x_i^{\alpha\dot{\alpha}}, \theta_i^{I\alpha}\}$. The \textit{on-shell space} is spanned by coordinates $\{\lambda_i^{\alpha}, \tilde{\lambda}_i^{\dot{\alpha}}, \eta_i^{I}\}$ and the \textit{dual space} is spanned by $\{\lambda_i^{{\alpha}},  x_i^{\alpha\dot{\alpha}}, \theta_i^{I\alpha}\}$ \cite{Drummond08}. All generators of any algebra defined on the full space must (super)commute  with \eqref{C_4d}  in order to preserve this subspace.  The generators of the dual superconformal algebra can be found to be \cite{Plefka09}:
\be
\label{G}
G_{4D}=\{P_{\alpha\dot{\alpha}}, M_{\alpha\beta}, \bar{M}_{\dot{\alpha}\dot{\beta}}, D, K^{\alpha\dot{\alpha}}, Q_{I\alpha}, \bar{Q}^I_{\dot{\alpha}}, S^{A}_{\alpha}, \bar{S}_{I\dot{\alpha}}, R^I{}_J, C\}~.
\ee
The most interesting are the non-local generators 
\be
\label{K4d}
K^{\alpha\dot{\alpha}}=\sum_i \{x^{\dot{\alpha}\epsilon}_i\theta^{I\alpha}_i\partial_{iI\epsilon}+ x^{\alpha\dot{\beta}}_ix^{\gamma \dot{\alpha}}_i\partial_{i\gamma \dot{\beta}}+x_{i}^{\alpha\dot{\epsilon}}\tilde{\lambda}^{\dot{\alpha}}_i\partial_{i\dot{\epsilon}}+x_{i+1}^{\epsilon\dot{\alpha}}\lambda^{\alpha}_i\partial_{i\epsilon} 
\ee
\begin{equation*}
~~~~~~~~+\tilde{\lambda}^{\dot{\alpha}}_i\theta^{I\alpha}_{i+1}\partial_{iI}\}~,
\end{equation*}
\be
\label{S4d}
S^{I\alpha}=\sum_i\{-\theta^{J\alpha}_i\theta^{I\beta}_i\partial_{iJ\beta}+x^{\alpha\dot{\gamma}}_i\theta^{I\beta}_i\partial_{i\beta\dot{\gamma}}+\theta_i^{I\beta}\lambda^{\alpha}_i\partial_{i\beta} 
\ee
\begin{equation*}
~~~~~~~~+x_{i+1}^{\alpha\dot{\beta}}\eta^I_i\partial_{i\dot{\beta}}+\eta^{I}_i\theta^{N\alpha}_{i+1}\partial_{iN}\}~.
\end{equation*}
(For conventions see appendix \ref{conventions}.) We can solve the constraints \eqref{C_4d} by
\be
x_i^{\alpha\dot{\alpha}}=x_1^{\alpha\dot{\alpha}}-\sum_{j=1}^{i-1}\lambda^{\alpha}_j\tilde{\lambda}^{\dot{\alpha}}_j~, \qquad
\theta^{I\alpha}_i=\theta^{I\alpha}_1-\sum_{j=1}^{i-1}\eta^I_j\lambda_j^{\alpha}~,
\ee
plug back into the all generators in \eqref{G} and compare with the superconformal generators \eqref{g_4D}. For more details of this procedure, see ref.\ \cite{Plefka09}. It turns out that most of the dual superconformal generators $G_{4D}$ after this procedure coincide with some of the superconformal generators $g_{4D}$. However \eqref{K4d} and \eqref{S4d} produce new bi-local expressions of so-called \textbf{level-1} generators. (We say that generators of superconformal algebra $g_{4D}$ are \textbf{level-0} generators.) \\
\indent Following the standard notation let's denote $J^{(0)}$ as generators of the standard superconformal algebra and $J^{(1)}$ as level-1 generators. It is possible to show that they satisfy \cite{Plefka09}, \cite{MacKay04}
\begin{equation}
\label{Yang_def}
\left[J^{(0)}_a,J^{(0)}_b\right]=f_{ab}{}^{c}J^{(0)}_c, \qquad \left[J^{(1)}_a, J^{(0)}_b\right]=f_{ab}{}^cJ^{(1)}_c~.
\end{equation}
The symbol
$
\left[\cdot,\cdot\right]
$ means supercommutator
$
\left[A,B\right]:= AB-(-1)^{|A||B|}BA
$.  The Grassmann parity $|A|$ is 0 if $A$ is bosonic and 1 if $A$ is fermionic. Moreover these generators satisfy Serre relations \cite{Drinfeld85}. For review see ref.\ \cite{MacKay04}. It is also possible to write generators  $J^{(1)}$ as a bi-local expression   
\be
J^{(1)}_a=f_a{}^{bc}\sum_{1\leq i< j\leq n}J^{(0)}_{ib}J^{(0)}_{jc}~
\ee
of $J^{(0)}$.
The fact that the level-0 and level-1 generators satisfy \eqref{Yang_def}  together with Serre relations give rise to an infinite tower of generators that span the so-called Yangian algebra.

\section{Construction of the 6D dual superconformal algebra} \label{6Dconstruction}
In analogy with 4D we start with the constraints. The constraints \eqref{C_4d} can be viewed as a solution to the constraints imposed by momentum and supermomentum conservation encoded in products of delta functions $\delta(p)\delta(q)$. Momentum conservation in the dual language is simply the identification $x_1 \equiv x_{n+1}$ (analogously $\theta_1 \equiv \theta_{n+1}$) which can be achieved by replacing the product of momentum and supermomentum delta functions by the product $\delta(x_1-x_{n+1})\delta(\theta_1-\theta_{n+1})$. Therefore we call the scattering amplitude on the dual space the following distribution
\be
\label{dual_amp}
\mathcal{A}=\delta\left(x_1-x_{n+1}\right)\delta\left(\theta_1-\theta_{n+1}\right)f_n\left(x_{ij}, \theta_{kl}, \lambda\right)~,
\ee
where $x_{ij}=x_i-x_j$, $\theta_{ij}= \theta_i-\theta_j$, and $\lambda$ represent spinor dependence. We can straightforwardly see that the scattering amplitudes are annihilated by following generators\footnote{Sums in generators goes from 1 to $n+1$ because we can treat $x_1$ and $x_{n+1}$ as independent. The identification $x_1\equiv x_{n+1}$ is imposed by $\delta(x_{1}-x_{n+1})$ in an amplitude.}
\be
P=\sum_{i=1}^{n+1}\frac{\partial}{\partial x_i}, \qquad Q=\sum_{i=1}^{n+1}\frac{\partial}{\partial \theta_i}~,
\ee
because the amplitude \eqref{dual_amp} depends only on the differences of $x_i$ or $\theta_i$. \\
\indent The discussion was independent of the spacetime dimension so far. Let's now specify the dimension of the spacetime to be $D=6$. Before constructing the algebra we will discuss the chirality. 6D vectors have two representations $p^{AB}_i$ and $\tilde{p}_{iAB}$. Each representation of momenta can be expressed in terms of spinors $\lambda^{Aa}_i$ and $\tilde{\lambda}_{iA\dot{a}}$. Each of the spinor variables $\lambda^{Aa}_i$ or $\tilde{\lambda}_{iA\dot{a}}$ carry one of the little group indices $a$ or $\dot{a}$, and each of $a$ or $\dot{a}$ belong to one of two distinct $SU(2)$ groups in the little group $SO(4)\cong [SU(2) \times SU(2)]/\mathbb{Z}_2$. Therefore we say we have both chiralities.  From lower-dimensional perspective all kinematical data are already present in $\lambda^{Aa}_i$ (see chapter \ref{DimRed}).  Moreover the  usage of the $\lambda^{Aa}_i$ only allows us to construct the chiral superconformal algebra like $\mathcal{N}=(2,0)$ whose dimensional reduction to 4D leads to $psu(2,2|4)$ \cite{Chern09}. Therefore we use the spinors $\lambda_i^{Aa}$ only. Consequently we parametrize the fermionic directions with Grassmann coordinates $\eta^{Ia}_i$, where the $R$-symmetry index $I$ runs from $1$ to $\mathcal{N}$, and  where $\mathcal{N}$ is the number of supersymmetries.

We will represent generators of the dual algebra in a space spanned by set of coordinates
\begin{equation}
\label{presuperspace}
\{\lambda_i^{Aa}, \eta_i^{Ia}, x^{AB}_i, \theta^{IA}_i\}~.
\end{equation}
(Note that this space \eqref{presuperspace} needs to be  extended later, cf.\ \eqref{superspace}.) The construction of the dual superconformal algebra is based on three guiding principles:
\begin{itemize}
\item The generators must commute with constraints 
\be
\label{constraints6D}
C_1^{EF} := x^{EF}_{i}-x^{EF}_{i+1}-\lambda_i^{Ea}\lambda_{ia}^F \approx 0, 
\ee
\begin{equation*}
C_2^{KF} := \theta_i^{KF}-\theta_{i+1}^{KF}-\eta_i^{Ka}\lambda_{ia}^F \approx 0~.
\end{equation*}
\item The generators form a superconformal algebra.
\item Ansatz for dual generators (in analogy with 4D and discussion above) is
\be
\label{ansatz}
P_{AB}=\sum_{i=1}^{n+1}\frac{\partial}{\partial x^{AB}_i}, \qquad Q_{IA}=\sum_{i=1}^{n+1}\frac{\partial}{\partial \theta^{IA}_i}~.
\ee
\end{itemize}

\subsection{Bosonic generators}
Let's start with the dilatation generator which measures dimension of momentum operator according to $[D, P^{AB}]=P^{AB}$. Therefore we can start with the ansatz\footnote{The word "$ \text{t.b.m.}$" is meant as a warning to the reader that the formula needs to be modified for various reasons explained in the text. t.b.m. = to be modified.}
\be
D=-\sum_{i}x^{AB}_i\partial_{iAB}~ +~ \text{t.b.m.}~ .
\ee
Requiring $D$ to commute with the constraints $C_1$ and $C_2$ leads to 
\be
D=-\sum_i \left\{x^{AB}_i\partial_{iAB}+\frac{1}{2}\lambda^{Aa}_i\partial_{iAa}+\frac{1}{2}\theta^{IA}_i\partial_{iIA}\right\}~ .
\ee
We can deduce the form of the Lorentz generators from it to be 
\be
M^A{}_{B}=\sum_i \left\{x^{AC}_i\partial_{iBC}+\frac{1}{2}\lambda^{Aa}_i\partial_{iBa}+\frac{1}{2}\theta^{IA}_i\partial_{iIB}\right\}~ +~ \text{t.b.m.}~.
\ee
The dilatation generator helps us to make the Lorentz generators traceless simply by adding $\frac{1}{4}D$ 
\be
M^A{}_{B}=\sum_i \left\{x^{AC}_i\partial_{iBC}+\frac{1}{2}\lambda^{Aa}_i\partial_{iBa}+\frac{1}{2}\theta^{IA}_i\partial_{iIB}+\frac{1}{4}\delta_{B}^{A}D\right\}~.
\ee
We have the following algebra
\begin{equation}
\left[D, P_{AB}\right]=P_{AB}~, \qquad \left[D, M^A{}_B\right]=0~, \qquad 
\end{equation}
\begin{equation*}
\left[P_{AB}, M^{E}{}_{F}\right]=\frac{1}{2}\delta_{[A}^{E}P_{FB]}-\frac{1}{4}\delta_{F}^{E}P_{AB} ~,
\end{equation*}
\begin{equation*}
\left[M^{A}{}_{B}, M^{E}{}_{F}\right]=\frac{1}{2}\delta_{B}^EM^{A}{}_F-\frac{1}{2}\delta_{F}^AM^{E}{}_{B}~.
\end{equation*}
We can now continue with the special conformal generator. It is possible to start with rules for conformal inversion given for example in \cite{Plefka14}. Here we will instead use an ansatz inspired by 3D \cite{Huang_Lipstein10} and 4D \cite{Drummond08} 
\be
\label{K_ansatz}
K^{AD}=-\sum_i x_i^{[AB}x_i^{CD]}\partial_{iBC} ~ +~ \text{t.b.m.}~.
\ee
Requiring commutation of \eqref{K_ansatz} with constraints \eqref{constraints6D} leads to
\begin{equation}
K^{AD}=\sum_i   \left\{x_i^{[AE}\theta^{MD]}_i\partial_{iME}-x^{[AB}_i x^{CD]}_i\partial_{iBC} \right. 
\end{equation}
\begin{equation*}
~~~~~~~~~~+\frac{1}{2}\left(x^{[AE}_i+ x^{[AE}_{i+1}\right)\lambda^{D]a}_i\partial_{iEa}+\left.\frac{1}{2}\lambda_i^{[Aa}\left(\theta_{i}^{MD]}+\theta_{i+1}^{MD]}\right)\partial_{iMa}\right\}~ 
\end{equation*}
\begin{equation*}
~~~~~~~~~~+~ \text{t.b.m.}~.
\end{equation*}

This determines the first few terms of the bosonic generators.

\subsection{Fermionic generators and new $y^{IJ}_i$ variables}
We want to deal with a superalgebra of the form
\be
\label{susy}
\left[Q_{IA}, \bar{Q}^J_B\right]=\delta_I^JP_{BA}~.
\ee
Every $R$-symmetry index $I,J, ....$ runs from $1$ to $\mathcal{N}$, where $\mathcal{N}$ is the number of supersymmetries which we keep general.

The formulation \eqref{susy} is chosen in order to make it as similar to 4D as possible because it is more suitable for superization, although these supercharges may not be the physical  6D supercharges $\mathcal{Q}_{IA}$. To make contact with the standard 6D $(2,0)$ formulation (e.g.\ in \cite{Huang_Lipstein10a}) of supercharges, we use an antisymmetric metric $\Omega_{IJ}$ of $USp(4)$ to lower the $R$-symmetry index on $\bar{Q}^J_A$. Next define 
\begin{equation}
\mathcal{Q}_{IA}=sQ_{IA}+\frac{\Omega_{IJ}}{2s}\bar{Q}^{J}_A ~,
\end{equation}
where $s$ is a free parameter. Then we have
\begin{equation}
\left[\mathcal{Q}_{IA}, \mathcal{Q}_{JB}\right]=\Omega_{IJ}P_{AB}~.
\end{equation}
Similarly we can define $\mathcal{S}_I^A=sS_I^A+\frac{\Omega_{IJ}}{2s}\bar{S}^{JA}$. 

Looking at equation \eqref{susy} it is clear that $\bar{Q}_B^J$ must be of the form
\be
\label{Qbar_ans}
\bar{Q}_B^J=\sum_i\theta^{JC}_i\partial_{iBC}~+~\text{t.b.m.}~. 
\ee
According to \eqref{ansatz}  $Q\sim\partial_{IA}$ and therefore  $\bar{Q}$ must contain $\theta$ contracted with an object with two spinor indices down. Requiring \eqref{Qbar_ans} to commute with $C_1$ from \eqref{constraints6D} leads to
\be
\label{Qbar}
\bar{Q}_B^J=\sum_{i}\left\{\theta^{JC}_i\partial_{iBC}+\frac{1}{2}\eta^{Ja}_i\partial_{iBa}\right\}~+~\text{t.b.m.}~.
\ee
We can compare the structure of \eqref{Qbar} with the 4D form of the corresponding generator for example in ref.\ \cite{Plefka09} and it agrees. However \eqref{Qbar} does not commute with constraint $C_2$ in \eqref{constraints6D}, which can be straightforwardly seen from 
\be
\left[\bar{Q}_B^J, C_2^{KF}\right] \propto \delta^F_B\eta^{Ja}_i\eta^K_{ia}~.
\ee 
This spoils the construction of $\bar{Q}$ because we cannot cancel this term by any combination of full-space coordinates. This problem is known and  appeared also in the construction of the dual superconformal symmetry in 3D for example in \cite{Huang_Lipstein10}. The solution is to introduce new coordinates $y^{IJ}_i$ related with $\eta^{Ia}_i$ via
\be
\label{C_3}
C_3^{KL} := y_i^{KL}-y_{i+1}^{KL}-\eta^{Ka}_i\eta^L_{ia} \approx 0~,
\ee
and with a bit of work, we find that
\be
\bar{Q}_B^J=\sum_i\left\{\theta^{JC}_i\partial_{iBC}+\frac{1}{2}\eta^{Ja}_i\partial_{iBa}-\frac{1}{2}y^{JK}_i\partial_{iKB}\right\}~ .
\ee
It is easy now to verify that the commutators of $\bar{Q}$  with all three constraints vanish modulo constraints.  

By introducing new coordinates $y^{IJ}$ we have to do some modifications of the previously said. First we extend the space of coordinates  \eqref{presuperspace}  to be
\begin{equation}
\label{superspace}
\mathcal{M}=\{\lambda_i^{Aa}, \eta_i^{Ia}, x^{AB}_i, \theta^{IA}_i,y^{IJ}_i\}~.
\end{equation}
The presence of $y^{IJ}$ variables is a new feature in 6D and differs from massless $D=4$ $\mathcal{N}=4$ SYM.  We will show in section \ref{DimRed} that the 6D model can be  consistently dimensionally reduced to the dual superconformal algebra of massless $D=4$ $\mathcal{N}=4$ SYM. Secondly, 
we add a new constraint so we must check and  potentially modify all generators, for example $K$ must be modified to the form
\begin{equation}
\label{K_6D}
K^{AD}=\sum_i   \left\{x_i^{[AE}\theta^{MD]}_i\partial_{iME}-x^{[AB}_i x^{CD]}_i\partial_{iBC}+\theta^{M[A}_i\theta^{ND]}_i\partial_{iMN} \right.
\end{equation}
\begin{equation*}
~~~~~~~~~~+\frac{1}{2}\left(x^{[AE}_i+ x^{[AE}_{i+1}\right)\lambda^{D]a}_i\partial_{iEa}+\left.\frac{1}{2}\lambda_i^{[Aa}\left(\theta_{i}^{MD]}+\theta_{i+1}^{MD]}\right)\partial_{iMa}\right\}~.
\end{equation*}
We can now use the commutator of $Q$ resp. $\bar{Q}$ with $K$ to obtain $S$ resp. $\bar{S}$. Explicitly
\be
\label{S_6D}
S^A {}_I=\sum_i\left\{x^{AB}_i\partial_{iIB}+\lambda^{Aa}_i\partial_{iIa}-2\theta^{JA}_i\partial_{iIJ}\right\}~,
\ee
\be\bar{S}^{IA}=\sum_i&\left\{\theta^{IB}_i\theta^{NA}_i\partial_{iNB}-2\theta^{IB}_ix^{CA}_i\partial_{iBC}\right.
\ee
\begin{equation*}
~~~~~~~~-2\theta^{UA}_iy^{IN}_i\partial_{iUN}+x^{AB}_iy^{IN}_i\partial_{iNB}
\end{equation*}
\begin{equation*}
~~~~~~~~-\frac{1}{2}\eta^{Ia}_i\left(x^{AB}_i+x^{AB}_{i+1}\right)\partial_{iBa}+\frac{1}{2}\eta^{Ia}\left(\theta_{i}^{NA}+\theta^{NA}_{i+1}\right)\partial_{iNa}
\end{equation*}
\begin{equation*}
~~~~~~~~+\frac{1}{2}\lambda^{Aa}\left(y_{i}^{IN}+y^{IN}_{i+1}\right)\partial_{iNa}+\left.\frac{1}{2}\lambda^{Aa}_i\left(\theta^{IB}_i+\theta_{i+1}^{IB}\right)\partial_{iBa}\right\}~.
\end{equation*}
So far we found several generators of the dual superconformal algebra, namely $P_{AB}$, $M^A{}_B$, $D$, $K^{AB}$, $Q_{IA}$, $\bar{Q}_A^{J}$, $S^{A}{}_I$ and $\bar{S}^{IA}$. Despite new terms containing $y$ the structure of most of them is as same as in $D=4 ~ \mathcal{N}=4$ SYM. However the structure of $K^{AB}$ and $\bar{S}^{IA}$ is different. This will be explained in section \ref{DimRed}.

\section{Superization} \label{Superization}
\indent It is possible to significantly simplify relatively long expressions for all the dual superconformal generators but mainly the $K$ generator \eqref{K_6D} and the $S$ generator \eqref{S_6D}. This simplification can be achieved with help of superindices $\mathcal{A}=(A,I)$ where the first entry corresponds  to the spinor index and the second entry corresponds to the $R$-symmetry index. For instance imagine  an object with two external indices which is composed out of the sum of terms with the structure of a product of two variables and one derivative (all internal indices contracted). If external indices are chosen to be spinor indices  we obtain similar expression to the \eqref{K_6D}. \\
\indent Motivated by the method described in the previous paragraph we can introduce new variables 
\be
\Lambda^{\mathcal{A}a}_i :=
\left(
\begin{array}{c}
\lambda^{Aa}_i\\
\eta^{Ia}_i
\end{array}
\right)~.
\ee
It is easy then to rewrite all constraints $C_1^{EF}$, $C_2^{KF}$ and $C_3^{KL}$ from \eqref{constraints6D} and \eqref{C_3} into the form
\be
\label{constraints_super}
X^{\mathcal{AB}}_i-X^{\mathcal{AB}}_{i+1}-\Lambda^{\mathcal{A}a}_i\Lambda^{\mathcal{B}}_{ia}\approx 0~.
\ee
Then $X^{\mathcal{AB}}_i$ is of the form
\be
X^{\mathcal{AB}}_i
=\left(
\begin{array}{cc}
x^{AB}_i& \theta^{AJ}_i\\
\theta^{IB}_i& y^{IJ}_i
\end{array}
\right)~.
\ee
It is now easy to see the symmetry of $X^{\mathcal{AB}}_i$ to be\footnote{This implies antisymmetry of $\theta^{IA}$ in $I$ and $A$, which superficially may seem strange because of different types of indices. However we can understand this from constraint $\theta^{IA}_i-\theta^{IA}_{i+1}\approx \eta^{Ia}_i\lambda^{A}_{ia}$ where the antisymmetry comes from $\epsilon_{ab}$ in little group contraction. If we replace $I$ and $A$ we have to also change the order of $\eta$ and $\lambda$ and raise and lower little group index which costs a sign.}
\be
X^{\mathcal{AB}}_i=(-1)^{|\mathcal{A}||\mathcal{B}|+1}X^{\mathcal{BA}}_i~.
\ee
The superspace \eqref{superspace} can now be  written as $\mathcal{M}=\{\Lambda^{\mathcal{A}a}_i, X^{\mathcal{AB}}_i\}$. Therefore we propose the following form of generators in superized form
\be
\label{P_super}
\mathds{P}_{\mathcal{AB}}=\sum_i\frac{\partial}{\partial X^{\mathcal{AB}}_i}~,
\ee
\be
\label{M_super}
\mathds{M}^{\mathcal{A}}{}_{\mathcal{B}}=\sum_i\left\{X^{\mathcal{AC}}_i\frac{\partial}{\partial X^{\mathcal{BC}}_i}+\frac{1}{2}\Lambda^{\mathcal{A}a}\frac{\p}{\p \Lambda^{\mathcal{B}a}_i}\right\}~,
\ee
\be
\label{D_super}
\mathds{D}=-\sum_i\left\{X^{\mathcal{AB}}_i\frac{\partial}{\partial X^{\mathcal{AB}}_i}+\frac{1}{2}\Lambda^{\mathcal{A}a}\frac{\p}{\p \Lambda^{\mathcal{A}a}_i}\right\}~,
\ee
\be
\label{K_super}
\mathds{K}^{\mathcal{AB}}=\sum_i \left\{\alpha_S(-1)^{|\mathcal{B}|(|\mathcal{C}|+|\mathcal{D}|)}X^{\mathcal{AC}}_iX^{\mathcal{DB}}_i\frac{\p}{\p X^{\mathcal{CD}}_i}\right.
\ee
\begin{equation*}
~~~~~~+\left.\beta_S(-1)^{|\mathcal{B}||\mathcal{C}|+1}\left(X^{\mathcal{AC}}_i+X^{\mathcal{AC}}_{i+1}\right)\Lambda^{\mathcal{B}a}_i\frac{\p}{\p \Lambda^{\mathcal{C}a}}_i\right\}\\
+(-1)^{|\mathcal{A}||\mathcal{B}|+1}\left(\mathcal{A}\leftrightarrow\mathcal{B}\right)~,
\end{equation*}

where $\alpha_S$ and $\beta_S$ are two new free constants for now. They are determined by requiring commutation of \eqref{K_super} with  \eqref{constraints_super}.
It is easy to see that \eqref{P_super}, \eqref{M_super} and  \eqref{D_super} commute with \eqref{constraints_super}. The remaining commutation of $\mathds{K}$ with constraint imposes the constraint on coefficients
\be
\alpha_S-2\beta_S=0~.
\ee
\indent All generators now commute with the constraint (modulo this constraint). Once we know the form of generators it is straightforward to find the algebra they satisfy. Only non-vanishing commutators are  
\be
\left[\mathds{D},\mathds{P}_{\mathcal{AB}}\right]=\mathds{P}_{\mathcal{AB}}~, \qquad \left[\mathds{P}_{\mathcal{CD}}, \mathds{M}^{\mathcal{A}}{}_{\mathcal{B}}\right]=\delta^{\mathcal{A}}_{\mathcal{C}}\mathds{P}_{\mathcal{BD}}+(-1)^{|\mathcal{D}||\mathcal{C}|+1}\delta^{\mathcal{A}}_{\mathcal{D}}\mathds{P}_{\mathcal{BC}}~,
\ee
\be
\left[\mathds{K}^{\mathcal{AB}}, \mathds{D}\right]=\mathds{K}^{\mathcal{AB}}~, \qquad
\left[\mathds{M}^{\mathcal{A}}{}_{\mathcal{B}},\mathds{K}^{\mathcal{EF}}\right]=\delta^{\mathcal{E}}_{\mathcal{B}}\mathds{K}^{\mathcal{AF}}+(-1)^{|\mathcal{E}||\mathcal{F}|+1}\delta^{\mathcal{F}}_{\mathcal{B}}\mathds{K}^{\mathcal{AE}}~,
\ee
\be
\left[\mathds{M}^{\mathcal{A}}{}_{\mathcal{B}},\mathds{M}^{\mathcal{E}}{}_{\mathcal{F}}\right]=\frac{1}{2}\delta^{\mathcal{E}}_{\mathcal{B}}\mathds{M}^{\mathcal{A}}{}_{\mathcal{F}}+(-1)^{(|\mathcal{A}|+|\mathcal{B}|)(|\mathcal{E}|+|\mathcal{F}|)+1}\delta^{\mathcal{A}}_{\mathcal{F}}\mathds{M}^{\mathcal{E}}{}_{\mathcal{B}}~,
\ee
\be
\left[\mathds{K}^{\mathcal{AB}},\mathds{P}_{\mathcal{EF}}\right]=\alpha_{S}\left[(-1)^{|\mathcal{B}||\mathcal{F}|+1}\delta_{\mathcal{E}}^{\mathcal{A}}\mathds{M}^{\mathcal{B}}{}_{\mathcal{F}}+(-1)^{|\mathcal{E||F}|+1}(\mathcal{E}\leftrightarrow\mathcal{F})\right]
\ee
\begin{equation*}
~~~~~~~~~~~~~~~~~~~+(-1)^{\mathcal{|A||B|}+1}(\mathcal{A}\leftrightarrow\mathcal{B})~.
\end{equation*}
The only nontrivial calculation is commutator $[\mathds{K},\mathds{K}]=0$. When calculating this commutator in non-superized form we have to use the Schouten identity twice. Similar argument must be used in calculating the superized version also, however it is necessary to use superized version of Schouten identity
\be
\sum_{\mathcal{A,B,C}~\mathrm{cycl}.} (-1)^{\mathcal{|A||C|}}\Lambda^{\mathcal{A}a}_i\Lambda^{\mathcal{B}}_{ia}\Lambda^{\mathcal{C}b}_i=0~.
\ee
It is easy to see that it holds simply by writing out all terms explicitly.\\
\indent Extracting all possible combination of indices leads to all possible generators. There is however one subtlety. The superized version of the dilatation generator is not what is  conventionally called the dilatation generator in 6D. An explicit formula for $\mathds{D}$ is
\begin{equation}
\label{D_super_exp}
\mathds{D}= -\sum_i\left\{X^{\mathcal{AB}}_i\frac{\partial}{\partial X^{\mathcal{AB}}_i}+\frac{1}{2}\Lambda^{\mathcal{A}a}\frac{\p}{\p \Lambda^{\mathcal{A}a}_i}\right\}
\end{equation}
\begin{equation}
~~=-\sum_i\left\{x^{AB}_i\partial_{iAB}+\theta^{IA}_i\partial_{iIA} +y^{IJ}_i\partial_{iIJ}+\frac{1}{2}\lambda^{Aa}_i\partial_{iAa}+\frac{1}{2}\eta^{Ia}_i\partial_{iIa}\right\}~.
\end{equation}
It is clear that \eqref{D_super_exp} contains what is usually called the dilatation generator in 6D. Therefore we can rewrite it as
\be
\mathds{D}=D-\frac{1}{2}\sum_{i}\left\{\eta^{Ia}_i\partial_{iIa}+ \theta^{IA}_i\partial_{iIA}+2y^{IJ}_i\partial_{iIJ}\right\}~.
\ee
The remaining expression can be associated with the so-called hypercharge $B$. Similar formulas for the hypercharge can be found for example in \cite{Plefka14} or in \cite{Drummond08}. The sum of dilatation generator and hypercharge was also mentioned in \cite{Bargheer11}.

\section{Dimensional reduction}\label{DimRed}

 Once we have found the generators of the dual superconformal algebra it is now straightforward to compare them with known examples. Such examples are massless $\mathcal{N}=4$ SYM theory in 4D and massless $\mathcal{N}=6$ ABJM theory in 3D. For this reason we perform dimensional reduction of the generators.  To obtain 4D massless theories  we choose $p_4=p_5=0$. Plugging into \eqref{Lambda_6D} we find how 4D spinors (vectors) sit inside 6D spinors (vectors):
\be
\label{4D_in_6D}
\lambda^{Aa}_i=\left(
\begin{array}{cc}
0&\lambda_{i\alpha}\\
\tilde{\lambda}^{\dot{a}}_i&0
\end{array}
\right)~, \qquad 
p_i^{AB}=
\left(
\begin{array}{cc}
0&-p_{i\alpha}^{\dot{\beta}}\\
p_{i\beta}^{\dot{\alpha}}&0
\end{array}
\right)~.
\ee
Similarly we can relate 6D and 3D spinors where the convention is to choose $p_5=p_4=p_2=0$. So we find
\be
\label{3D_in_6D}
\lambda^{Aa}_i=\left(
\begin{array}{cc}
0&\lambda_{i\alpha}\\
\lambda^a_i&0
\end{array}
\right)~, \qquad 
p_i^{AB}= \left(
\begin{array}{cc}
0 & -p_{i\alpha}^{\beta}\\
p^{\alpha}_{i\beta} & 0
\end{array}
\right)~. 
\ee 
To dimensionally reduce all generators we have to specify the form of $\eta^{Ia}$ also. However this is different for $\mathcal{N}=4$ SYM and $\mathcal{N}=6$ ABJM  for starters because they have different number of supersymmetries. Performing the dimensional reduction we have to keep in mind the right number of supersymmetries. The Grassmann variables $\eta^{Ia}_i$ have naively $8$ degrees of freedom for $\mathcal{N}=4$. However this is twice the maximum number of on-shell SUSY variables. This issue is also discussed e.g.\ in \cite{Huang_Lipstein10a}, where the unphysical degrees of freedom are removed by introducing harmonic coordinates (which break the $R$-symmetry), or by choosing the physical degrees of freedom according to their $SU(2)$ weights, e.g. $\eta^{I+}_i$ (which break the $SU(2)$ little group). We simply puts the right number of $\eta$'s in the matrix form $\eta^{Ia}_i$ and the rest are put to 0.

\subsection{4D massless $\mathcal{N}=4$ SYM}\label{DimReduction4D}
To dimensionally reduce to 4D we use \eqref{4D_in_6D}. What is left is the form of $\eta^{Ia}_i$. The supersymmetry of $\mathcal{N}=4$ SYM can be parametrized by (apart from the bosonic variables) the Grassmann-odd variables $\eta^{I}_i$ where the index $I$ runs from 1 to 4. According to \cite{Plefka09} the helicity generator for $\mathcal{N}=4$ SYM takes the form
\be
h_i=-\frac{1}{2}\lambda^{\alpha}_i\frac{\partial}{\partial{\lambda^{\alpha}_i}}+\frac{1}{2}\tilde{\lambda}^{\dot{\alpha}}_i\frac{\partial}{\partial{\tilde{\lambda}^{\dot{\alpha}}_i}}+\frac{1}{2}\eta^I_i\frac{\partial}{\partial \eta^{I}_i}~.
\ee
In other words  the 4D variables scale like $\lambda~\rightarrow~t\lambda$, $\tilde{\lambda}~\rightarrow~t^{-1}\tilde{\lambda}$  and $\eta~\rightarrow~t^{-1}\eta$ under little group scaling.  We can see that $\eta$  and $\tilde{\lambda}$ scale in the same way and therefore we propose 6D $\eta^{Ia}_i$ to be of the form
\be
\label{eta_4D}
\eta^{Ia}_i=\left(
\begin{array}{cc}
\eta^1_i & 0\\
\eta^2_i & 0\\
\eta^3_i & 0\\
\eta^4_i & 0
\end{array}
\right)~.
\ee
The choice \eqref{eta_4D} of $\eta^{Ia}_i$ also reflects the fact that $\eta^{Ia}_i$ has in general 8 components, however the physical relevant theory only needs half of them.  There are essentially two different ways how to solve this issue \cite{Huang_Lipstein10a}. We can either use (i) harmonic variables, discussed  e.g. in \cite{Ferrara_2000}, which breaks the $R$-symmetry, or we  can use (ii) $\eta^{I1}_i$ as our physical degrees of freedom, which breaks the $SU(2)$  little group symmetry. Our choice \eqref{eta_4D} is the second case (ii), because we want to keep the $R$-symmetry unbroken while the little group become $U(1)$ for 4D $\mathcal{N}=4$ SYM. Based on this we can then define the massless 4D $\mathcal{N}=4$ subspace of $\mathcal{M}$
\begin{equation}
\lambda^{Aa}_i=\left(
\begin{array}{cc}
0&\lambda_{i\alpha}\\
\tilde{\lambda}^{\dot{a}}_i&0
\end{array}
\right)~, \qquad 
\eta^{Ia}_i=\left(
\begin{array}{cc}
\eta^I_i & 0
\end{array}
\right)~.
\end{equation}
\indent
The first non-trivial check is 6D supertranslation generator $q^{IA}$
\be
q^{IA}_i=\eta^{Ia}_i\lambda_{ia}^A =\left(
\begin{array}{cc}
\eta^I_i\lambda_{i\alpha} & 0
\end{array}
\right)
\ee
which indeed reduces to the known form of the 4D generator. A much more interesting question is the presence of the $y^{IJ}_i$ variables which are related to the $\eta^{Ia}_i\eta^J_{ia}$ little group contraction via \eqref{C_3}. Plugging the \eqref{eta_4D} into \eqref{C_3} we find that this contraction is $0$ and therefore all $y_i$'s are identified ($y_{i}\approx y_{i+1}$ for all $1\leq i\leq n$). At this point we can get rid of all terms containing $y^{IJ}_i$. A good example is the generator $\bar{Q}$:
\be
\label{eq0}
\bar{Q}_B^J=\sum_i\left\{\theta^{JC}_i\partial_{iBC}+\frac{1}{2}\eta^{Ja}_i\partial_{iBa}-\frac{1}{2}y^{JK}_i\partial_{iKB}\right\}~,
\ee 
where identifying all $y_i$§'s leaves the last term to be
\be 
\label{eq1}
\bar{Q}^J_B \quad \rightarrow\quad  \ldots \quad -\frac{1}{2}y_1^{JK}\sum_i\partial_{iKB}= \quad \ldots \quad -\frac{1}{2}y_{1}^{JK}Q_{KB}~,
\ee
where ellipsis "$\ldots$" denote the untouched terms in \eqref{eq0}. We choose $y_1$ to be the 'surviving' variable. However, since all $y_i$'s are identified, $i$ can be arbitrary (from $1$ to $n$). The last term in \eqref{eq1} annihilates the  scattering amplitudes. Therefore we restrict (in this subsection) the space of objects, that our generators act on, to the scattering amplitudes (i.e.\ the generators annihilate them). As a result we can ignore the last term. Eq.\ \eqref{eq0} effectively remains the form of the $\bar{Q}$ generator
\be
\bar{Q}_B^J \quad\rightarrow \quad \sum_i\left\{\theta^{JC}_i\partial_{iBC}+\frac{1}{2}\eta^{Ja}_i\partial_{iBa}\right\}
\ee
whose dimensional reduction gives the right form of 4D generator.
\\
\indent The remaining dependence of generators on $y_i$'s is through derivatives. In Appendix  \ref{appdiracderiv} we show that terms involving derivatives with respect to the $y_i$'s can be neglected because on the subspace of $\mathcal{N}=4$ SYM all the $y_i$-derivatives vanish. 
This can be demonstrated on the $S$ generator
\be
S^A {}_I=\sum_i\left\{x^{AB}_i\partial_{iIB}+\lambda^{Aa}_i\partial_{iIa}-2\theta^{JA}_i\partial_{iIJ}\right\} \quad 
\ee
\begin{equation*}
~~~~\rightarrow  \sum_i\left\{x^{AB}_i\partial_{iIB}+\lambda^{Aa}_i\partial_{iIa}\right\}~,
\end{equation*}
where the first two terms agree with the known form of the $psu(2,2|4)$ generators and the last term vanishes because all the $y_i$-derivatives vanish. 
\\ \indent We can now dimensionally reduce the $K$ generator: 
\begin{equation*}
K^{AD}=\sum_i\left\{ x_i^{[AE}\theta^{MD]}_i\partial_{iME}- x^{[AB}_i x^{CD]_i}\partial_{iBC}+\frac{1}{2} x^{[AE}_i\lambda^{D]a}_i\partial_{iEa}\right.~~~
\end{equation*}
\begin{equation*}
~~~~~~~~~~\left. +\frac{1}{2}  x^{[AE}_{i+1}\lambda^{D]a}_i\partial_{iEa} +\frac{1}{2} \lambda_i^{[Aa}\theta_{i}^{MD]}\partial_{iMa}+\frac{1}{2}  \lambda_i^{[Aa}\theta_{i+1}^{MD]}\partial_{iMa}\right.
\end{equation*}
\begin{equation*}
~~~~~~~~~~\left. + \theta_{i}^{M[A}\theta_i^{ND]}\partial_{iMN}\right\} \rightarrow\left(
\begin{array}{cc}
0 &- K^{\dot{\delta}}{}_{\alpha}\\
 K^{\dot{\alpha}}{}_{\delta}& 0
\end{array}
\right)~,
\end{equation*}
where
\begin{equation}
\label{K_1}
K^{\dot{\delta}}{}_{\alpha}=\sum_i\left\{x^{\dot{\delta}\epsilon}_i\theta^I_{i\alpha}\partial_{iI\epsilon}+ x_{i\alpha}^{\dot{\beta}}x^{\gamma \dot{\delta}}_i\partial_{i\gamma \dot{\beta}}+\frac{1}{2} \left(x_{i}+x_{i+1}\right)_{\alpha}^{\dot{\epsilon}}\tilde{\lambda}^{\dot{\delta}}_i\partial_{i\dot{\epsilon}}
\right.
\end{equation}
\begin{equation*}
~~~~~~~~~+\frac{1}{2} \left(x_{i}+x_{i+1}\right)^{\epsilon\dot{\delta}}\lambda_{i\alpha}\partial_{i\epsilon} +\left.\frac{1}{2} \tilde{\lambda}^{\dot{\delta}}_i\left(\theta_i+\theta_{i+1}\right)^I_{\alpha}\partial_{iI}\right\}~.
\end{equation*}
This is not of the form that we would expect. To get the correct expression we need to get rid of half of the terms in last three parentheses. This can be achieved by adding an expression
\begin{equation*}
\sum_i\left\{-x_i^{\dot{\epsilon}}{}_{\alpha}\tilde{\lambda}^{\dot{\delta}}_i\partial_{i\dot{\epsilon}}-x_{i+1}^{\epsilon\dot{\delta}}\lambda_{i\alpha}\partial_{i\epsilon}+\theta_{i+1}^{I}{}_{\alpha}\tilde{\lambda}^{\dot{\delta}}_i\partial_{I}+x_{i+1}^{\dot{\epsilon}}{}_{\alpha}\tilde{\lambda}^{\dot{\delta}}_i\partial_{i\dot{\epsilon}}\right.
\end{equation*}
\begin{equation*}
+\left.x_{i}^{\epsilon\dot{\delta}}\lambda_{i\alpha}\partial_{i\epsilon}-\theta_{i}^{I}{}_{\alpha}\tilde{\lambda}^{\dot{\delta}}_i\partial_{iI}\right\} \approx\sum_i\left\{\lambda_{i\alpha}\tilde{\lambda}^{\dot{\delta}}_i\left(-\tilde{\lambda}^{\dot{\epsilon}}_i\partial_ {i\dot{\epsilon}}+\lambda^{\epsilon}_i\partial_{i\epsilon}-\eta^I_i\partial_{iI}\right)\right\}
\end{equation*}
\begin{equation*}
=\sum_i\left\{2p_{i\alpha}^{\dot{\delta}}c_i-2p_{i\alpha}^{\dot{\delta}}\right\}=-\sum_i2p_{i\alpha}^{\dot{\delta}}C_i~.
\end{equation*}

We are adding only terms that annihilate the scattering amplitudes of $\mathcal{N}=4$ SYM.  The central charge density commutes with all generators of the algebra and annihilates the amplitudes by itself. Next we add momentum, which is an obvious symmetry of scattering amplitudes. It is easy to see that commutators involving $K$ almost does not change  (by $K$ we mean the expression \eqref{K_1}) and it leads to change of $\bar{S}$ (cf.\ the  $[K^{\dot{\alpha}}_{i\alpha}, \bar{Q}_{i\dot{\beta}}^{I}]$ commutator below):

\begin{itemize}
\renewcommand{\labelitemii}{$\circ$}
\item $[K^{\dot{\alpha}}_{i\alpha}+(\mathrm{c.e.})\cdot p^{\dot{\alpha}}_{i\alpha},K^{\dot{\beta}}_{i\beta}+(\mathrm{c.e.})\cdot p^{\dot{\beta}}_{i\beta} ]=[K^{\dot{\alpha}}_{i\alpha},K^{\dot{\beta}}_{i\beta} ]$~,
\item $[K^{\dot{\alpha}}_{i\alpha}+(\mathrm{c.e.})\cdot p^{\dot{\alpha}}_{i\alpha}, P^{\dot{\beta}}_{i\beta}]=[K^{\dot{\alpha}}_{i\alpha}, P^{\dot{\beta}}_{i\beta}]$~,
\item $[K^{\dot{\alpha}}_{i\alpha}+(\mathrm{c.e.})\cdot p^{\dot{\alpha}}_{i\alpha}, Q_{iI\beta}]=[K^{\dot{\alpha}}_{i\alpha}, Q_{iI\beta}]$~,
\item $[K^{\dot{\alpha}}_{i\alpha}+(\mathrm{c.e.})\cdot p^{\dot{\alpha}}_{i\alpha}, \bar{Q}_{i\dot{\beta}}^{I}]=[K^{\dot{\alpha}}_{i\alpha},  \bar{Q}_{i\dot{\beta}}^{I}]-(\mathrm{c.e.})\cdot \delta_{\dot{\beta}}^{\dot{\alpha}}q_{i\alpha}^I$~.
\end{itemize}
Where $(\mathrm{c.e.})$ means a central element (such as numbers, central charges, ...).\\
\indent Next is the fermionic generator $\bar{S}$. Let's focus on the $y_i^{IJ}$ variables. Collecting everything that contain $y_i$'s we find
\begin{equation}
\bar{S}^{IA}= \ldots +\sum_i\left\{-2\theta^{UA}_iy^{IN}_i\partial_{iUN}+x^{AB}_iy^{IN}_i\partial_{iNB}\right.
\end{equation}
\begin{equation*}
~~~~~~~~~~~~\left. +\frac{1}{2}\lambda^{Aa}\left(y_{i}^{IN}+y^{IN}_{i+1}\right)\partial_{iNa}\right\}~.
\end{equation*}
Identifying all $y_i$'s we get an expression for the $S$ generator which annihilates the amplitudes by assumption. Therefore we can neglect these terms.  The rest is very similar to the $K$ story. To obtain the right form of the $\bar{S}$ generators we have to add  $\sum_i\left\{2q_{i\alpha}^{I}c_i-2q_{i\alpha}^{I}\right\}$. We can again check that all the (super)commutation relations remain the same 
\begin{itemize}
\renewcommand{\labelitemii}{$\circ$}
\item $[S^I_{i\alpha}+(\mathrm{c.e.})\cdot q^I_{i\alpha}, S^{J}_{i\beta}+(\mathrm{c.e.})\cdot q^J_{i\beta}]=[S^I_{i\alpha}, S^{J}_{i\beta}]$ ~,
\item $[S^I_{i\alpha}+(\mathrm{c.e.})\cdot q^I_{i\alpha}, p^{\dot{\beta}}_{\beta}]=[S^I_{i\alpha}, p^{\dot{\beta}}_{\beta}]$~,
\item $[S^I_{i\alpha}+(\mathrm{c.e.})\cdot q^I_{i\alpha}, q^{J}_{\beta}]=[S^I_{i\alpha},q^{J}_{\beta}]$~,
\item $[S^I_{i\alpha}+(\mathrm{c.e.})\cdot q^I_{i\alpha}, \bar{S}_{iJ\dot{\beta}}]=[S^I_{i\alpha}, \bar{S}_{iJ\dot{\beta}}]-(\mathrm{c.e.})\cdot \delta^I_{J}p_{i\alpha\dot{\beta}}$~.
\end{itemize}
\indent From \eqref{K_super} we can get generator with two $R$-symmetry indices. This object can be identified with the dual $R$-symmetry generator
\begin{equation}
\label{R_6D}
R^{IJ}=\sum_i\left\{y_i^{IK}y_i^{JL}\partial_{iKL}+\frac{1}{2}y_i^{\{IK}\theta_i^{J\}D}\partial_{iKD}- \theta_i^{\{IC}\theta_i^{J\}D}\partial_{iCD}
\right.
\end{equation}
\begin{equation*}
~~~~~~~~ -\frac{1}{4}\left(\theta_i^{\{IC}+\theta_{i+1}^{\{IC}\right)\eta_i^{J\}a}\partial_{iCa}+\left.\frac{1}{4}\left(y_i^{\{IK}+y_{i+1}^{\{IK}\right)\eta_i^{J\}a}\partial_{iKa}\right\}~.
\end{equation*}
On the subspace corresponding to the $\mathcal{N}=4$ SYM all the $y_i$'s are identified and therefore the first term in \eqref{R_6D} is proportional to the $R_{IJ}$ generator that annihilates the amplitudes by assumption. The third and the fourth term vanish due to dimensional reduction. What remains is proportional to the 4D non-traceless dual $R$-symmetry generator
\begin{equation}
R^{IJ} \rightarrow \frac{1}{2}y_1^{\{IK}\sum_i\left[\theta^{J\}\alpha}_i\partial_{iK\alpha}+\eta^{J\}}_i\partial_{iK}\right]~.
\end{equation}
\indent This finishes the dimensional reduction to massless $\mathcal{N}=4$ $D=4$ space. We found that most of the generators of the dual superconformal symmetry $psu(2,2|4)$ can be obtained from 6D dual superconformal algebra. Similar relation was found for superconformal symmetries also, e.g. in  \cite{Chern09} it was mentioned how the $psu(2,2|4)$ $\subset$ $psl(4|4, \mathbb{C})$ sits in $osp(8|4,\mathbb{C})$.  Exceptions are the Lorentz generators and the $R$-symmetry generators.  Dimensional reduction of the dual Lorentz generator gives non-symmetric version of 4D dual Lorentz generator and dimensional reduction of the dual $R$-symmetry generator gives the non-traceless version of 4D dual $R$-symmetry generator.

\subsection{3D $\mathcal{N}=6$ super Chern-Simons theory (ABJM theory)}
Another very interesting theory that has dual superconformal symmetry at tree-level \cite{Huang_Lipstein10},\cite{Lipstein13}\footnote{As proven in \cite{Lipstein13} the dual superconformal symmetry holds for tree-level amplitudes, however loop amplitudes are dual conformal covariant.} (and Yangian symmetry) is $\mathcal{N}=6$ ABJM theory in 3D. Looking at the structure of the algebra and structure of generators of this algebra (given for example in \cite{Huang_Lipstein10}) we can see that they have the same structure as the dual superconformal algebra in 6D. To compare it we have to specify the special form of Grassmann variable $\eta^{Ia}_i$ to be
\begin{equation}
\eta^{Ia}_i=\left(
\begin{array}{cc}
\eta^1_i & 0\\
\eta^2_i & 0\\
\eta^3_i & 0\\
0 & \eta^1_i \\
0 & \eta^2_i \\
0 & \eta^3_i 
\end{array}
\right)=\left(
\begin{array}{cc}
\eta^I_i & 0\\
0 & \eta^{I^{\prime}}_i 
\end{array}
\right).
\end{equation}
The 6D $R$-symmetry index $I$ is split into two indices $(I,I^{\prime})$ by dimensional reduction. Formally there is no difference between $I$ and $I^{\prime}$ but it is useful to distinguish them. This is similar to having lower  and upper spinor indices $\alpha$  in 3D. Little group contraction of two $\eta_i$'s gives us the correct form of $y_i$ variables because in 3D $y_{i,3D}^{IJ^{\prime}}$ is antisymmetric in $I$ and $J^{\prime}$ (in constrast to 6D where $y_i^{IJ}$ is symmetric in $I$ and $J$. With our conventions we have
\begin{equation}
y^{IJ}_{i,6D}=\left(
\begin{array}{cc}
0& y^{IJ^{\prime}}_{i,3D}\\
-y^{I^{\prime}J}_{i,3D}& 0
\end{array}
\right)~.
\end{equation}
The massless 3D $\mathcal{N}=6$ subspace of $\mathcal{M}$ is defined as
\begin{equation}
\lambda^{Aa}_i=\left(
\begin{array}{cc}
0&\lambda_{i\alpha}\\
\lambda^a_i&0
\end{array}
\right)~, \qquad 
\eta^{Ia}_i=\left(
\begin{array}{cc}
\eta^I_i & 0\\
0 & \eta^{I^{\prime}}_i 
\end{array}
\right).
\end{equation}

\indent The most interesting is to dimensionally reduce the generator that we obtain if we choose $\mathcal{A}$ and $\mathcal{B}$ in \eqref{K_super} to be $R$-symmetry indices  (for example $\mathcal{A}=I$ and $\mathcal{B}=J$).  
The dimensional reduction of \eqref{R_6D} to 3D leads to the expression
\begin{equation}
R^{IJ^{\prime}}=\sum_i \left\{
y^{IK^{\prime}}_iy^{J^{\prime}L}_i\partial_{iK^{\prime}L}+\frac{1}{2}y^{[IK^{\prime}}_i\theta_i^{J^{\prime}]\delta}\partial_{iK^{\prime}\delta}-\theta^{[I\gamma}_i\theta^{J^{\prime}]\delta}_i\partial_{i\gamma \delta}\right.
\end{equation}
\begin{equation*}
~~~~~~~~~-\frac{1}{4}\left(\theta^{[I\gamma}_i+\theta^{[I\gamma}_{i+1}\right)\eta_i^{J^{\prime}]}\partial_{i\gamma}+\left.\frac{1}{4}\left(y_i^{[IK^{\prime}}+y_{i+1}^{[IK^{\prime}}\right)\eta_i^{J^{\prime}]}\partial_{iK^{\prime}}
\right\}~,
\end{equation*}
which agrees with the dual $R$-symmetry generator in the appendix of \cite{Huang_Lipstein10}. Generator with similar structure was found also in \cite{Huang11}. All remaining generators of $osp(6|4)$ can be be derived in a similar way.

\section{6D analogue of 4D central charge?}

The main  inspiration for 6D dual superconformal algebra was the 4D dual superconformal symmetry of $\mathcal{N}=4$ SYM which is $psu(2,2|4)$.  This algebra contains the central charge
\begin{equation}
\label{central_charge}
C=\frac{1}{2}\sum_i\{-\lambda_i^{\alpha}\partial_{i\alpha}+\tilde{\lambda}_i^{\dot{\alpha}}\partial_{i\dot{\alpha}}+\eta_i^I\partial_{iI}\}~.
\end{equation}
We can enhance this generator to 6D. We can deduce the form of such generator from the 4D structure. We see from \eqref{central_charge} that such generator must have all Lorentz and $R$-symmetry indices contracted. It does not contain dual variables $x^{AB}$,  $y^{IJ}$ or $\theta^{IA}$. Therefore it must contain spinors $\Lambda^{\mathcal{A}}_{a}$ only. Such generator must have free little group indices, because contracting the little group indices we would get dilaton generator. Therefore we can define
\begin{equation}
\label{Center}
h_{ab}=\Lambda^{\mathcal{A}}_{(a}\partial_{\mathcal{A}b)}~,
\end{equation}
where () means symmetrization, and we can show that
\begin{itemize}
\item $\left[h_{ab}, \Lambda^{\mathcal{E}c}_i\Lambda^{\mathcal{F}}_{ic}\right]=0$~,
\item $\left[h_{ab}, \frac{1}{2}\Lambda^{\mathcal{E}c}_i\partial_{i\mathcal{E}c}+\alpha\right]=0$~,
\item $\left[h_{ab}, \Lambda^{\mathcal{E}c}_i\partial_{i\mathcal{F}c}\right]=0$~,
\item $\left[h_{ab}, \partial_{i\mathcal{E}c}\partial^{c}_{i\mathcal{F}}\right]=0$~.
\end{itemize}
We can see that $h_{ab}$ commutes with the superconformal algebra ($\Lambda^{\mathcal{E}a}_i\Lambda^{\mathcal{F}}_{ia}$ captures $p_i^{AB}$, $q_i^{IA}$ and $y_i^{IJ}$, and similarly for the rest).
The first commutator tells us that the generator \eqref{Center} commutes also with the dual constraint \eqref{constraints_super} (and therefore it is a well-defined operator in the full space also). We can see that the generator \eqref{Center} immediately commutes with $\mathbb{P}_{\mathcal{EF}},~\mathbb{D}, ~\mathbb{M}^{\mathcal{E}}{}_{\mathcal{F}}$. The only remaining nontrivial check is the commutator with $\mathds{K}^{\mathcal{EF}}$, and with a little bit of work, we find that 
\begin{itemize}
\item $\left[h_{ab}, \mathbb{K}^{\mathcal{EF}}\right]=0$~.
\end{itemize}

However $h_{ab}$ itself  is not  central, because it does not commute with itself\footnote{By () and $\langle\rangle$ parentheses we mean symmetrization.}
\begin{equation}
\label{algebra_321}
\left[h_{ab},h_{ef}\right]~=~2~\epsilon_{(e\langle a}h_{b\rangle f)}~,
\end{equation}
but merely form an $su(2)$ algebra. The Casimir operator of the $su(2)$ algebra \eqref{algebra_321} is
$h_{ab}h^{ab}~.$
Because the $h_{ab}$ commutes with all the generators of the symmetry algebra and  the Casimir operator commutes with all the $h_{ab}$ we can identify the \textit{central charge} to be proportional to the $su(2)$ Casimir operator
\begin{equation}
C~\propto~h_{ab}h^{ab}~.
\end{equation}
The full algebra is a direct sum
\begin{equation}
\mathcal{A}=\text{DSCA} \oplus su(2)~
\end{equation}
of the dual superconformal algebra and the $su(2)$ algebra.

\section{Example: amplitude toy model}
\label{examples}
As an example let us search for four-point and six-point functions that are manifestly invariant under the action of 6D dual super conformal generators \footnote{We ignore  for simplicity the generators $R^{IJ}$ and $R^I{}_J$  in this section \ref{examples}.}
 $P_{AB}$, $M^A{}_B$, $D^{\prime}$, $R_{IJ}$, $\bar{Q}^{I}_A$, $Q_{IA}$ and $K^{AB}$.  The strategy will be to study conformal inversion properties of individual parts of the functions inspired by scattering amplitudes and determine the remaining function, such that the whole object will be invariant under the action of generators listed above.

Conformal inversion rules were discussed e.g. in \cite{Dennen_10}, \cite{Plefka14} and \cite{Lipstein16} for 6D and \cite{Lipstein13} for 3D theories. Based on the literature, we use the following 6D dual conformal inversion rules
\begin{equation}
    \label{inversion_rules}
        I[x^{AB}_i]~=~\frac{x_{iAB}}{x^{2}_i }~=~\left(x^{-1}_i\right)_{AB}, \qquad I[\lambda_i^{Aa}]~=\frac{x_{iAB}\lambda^{B}_{ia}}{\sqrt{x_i^2x_{i+1}^2}}~, 
\end{equation}
\begin{eqnarray*}
    I[\theta^{IA}_i]~=~\frac{x_{iAB}\theta^{IB}_i}{x_i^2}, \quad
    I[\eta_i^{Ia}]~=~\sqrt{\frac{x^2_i}{x^2_{i+1}}}\left(\eta^{I}_{ia}-\frac{x_{iCD}\theta^{IC}_i\lambda^{D}_{ia}}{x_i^2}\right)~,
\end{eqnarray*}
\begin{eqnarray*}
         I[y^{IJ}_i]~=~y^{IJ}_i+\frac{x_{iAB}\theta^{IA}_i\theta^{JB}_i}{x^2_i}~.
\end{eqnarray*}
It is straightforward to use rules \eqref{inversion_rules} to obtain the inversion properties of basic building blocks, i.e.\  Lorentz invariants:
\begin{equation}
   I\left[ \left(x_{ij}\right)^2\right] ~=~ \frac{\left(x_{ij}\right)^2}{x^2_ix^2_j}~, \qquad x_{ij}~:=~x_i-x_j~, 
\end{equation}
\begin{equation*}
x_i^2~:=~-\frac{1}{8} \epsilon_{ABCD}x^{AB}_i x_{i}^{CD}~.
\end{equation*}
Inspired by scattering amplitudes, we also assume that the $n$-point functions we are looking for contain momentum and supermomentum conserving delta functions. This leads to the following ansatz in dual coordinates
\begin{equation}
    \mathcal{A}_n ~=~ \delta^{6}\left(x_1-x_{n+1}\right)\delta^8\left(\theta_1-\theta_{n+1}\right)f_n\left(x_{ij}, \theta_{ij}, y_{ij}\right)~,
\end{equation}
where we already used the requirement of invariance under $P_{AB}$, $Q_{IA}$ and $R_{IJ}$. From this invariance follows immediately that the unknown function $f_n$ can only depend on the differences of variables, e.g.\ $x_{ij} ~=~ x_i-x_j$, $\theta_{ij}~=~\theta_i-\theta_j$ and $y_{ij}~=~y_i-y_j$, where we omit spinor and fermionic indices for clarity. 
It is straightforward to find how the delta function product transforms under conformal inversion:
\begin{equation}
\label{delta_inversion}
    I[\delta^{6}\left(x_1-x_{n+1}\right)\delta^8\left(\theta_1-\theta_{n+1}\right)]~=~\left(x_1^2\right)^2\delta^{6}\left(x_1-x_{n+1}\right)\delta^8\left(\theta_1-\theta_{n+1}\right).
\end{equation}
Looking at \eqref{delta_inversion} we see that the product of delta functions has not vanishing dual conformal weight, as opposed to 4D $\mathcal{N}=4$ SYM \cite{Drummond08}. This is due to a mismatch in dimensions. 

Let us now be more specific. We focus on specific subset of invariants by using the ansatz
\begin{equation}
   \label{ansatzII}
    \mathcal{A}_n ~=~ \delta^{6}\left(x_1-x_{n+1}\right)\delta^8\left(\theta_1-\theta_{n+1}\right)\left[\left(y_1-y_{n+1}\right)^2\right]^{n-2}g_n\left(x_{ij}\right).
\end{equation}
The motivation for such an ansatz is following. The ansatz \eqref{ansatzII} contains standard delta functions. The function $f\left(x_{kl}, \theta_{kl}, y_{kl}\right)$ is chosen such that the full object is a top-form, i.e.\ it has maximal Grassmann degree. This is the purpose of the $[\left(y_1-y_{n+1}\right)^2]^{n-2}$, which is moreover invariant under the conformal inversion on the support of the delta functions.

This imply that the remaining function $g_n$ to be determined must be a function of $x_{kl}$ only. We assume that the dual inversion property of amplitude to be of general form
\begin{equation}
    \label{amplitude_inversion}
        I[\mathcal{A}_n]~=~\left[\displaystyle\prod_{i=1}^{n}\left(x_i^2\right)^{\omega}\right]\mathcal{A}_n~, \qquad \omega~\in~\mathbb{R}~.
\end{equation}
The relation \eqref{amplitude_inversion} immediately fixes the dual property of the function $g_n\left(x_{kl}\right)$
\begin{equation}
    \label{g_inversion}
        I[g_n\left(x_{ij}\right)]~=~\frac{\left(x_1^2 \cdot x_2^2 \cdot \ldots \cdot x^2_n\right)^{\omega}}{\left(x_1^2\right)^2}g_n\left(x_{ij}\right)~.
\end{equation}
Because the function $g_n\left(x_{ij}\right)$ depends on $x_{kl}$ only, \eqref{g_inversion}  fixes the dilaton weight of $g_n$ to be
\begin{equation}
        D~g_n\left(x_{ij}\right)~=~n\omega-2~.
\end{equation}
This agrees with the algebra modifications coming from requirement of invariance. The action of $K^{AB}$ on the \eqref{amplitude_inversion} tells us that $\mathcal{A}_n$ is not invariant but rather covariant under the action of $K^{AB}$
\begin{equation}
        K \equiv I~P~I\quad \Rightarrow\quad K^{AB}\mathcal{A}_n~=~-\frac{\omega}{2}\left(\sum_{i=1}^n x_i^{AB}\right)\mathcal{A}_n~.
\end{equation}
This motivates us to define $K^{\prime AB}$ such that it annihilates the amplitude,
\begin{equation}
        K^{\prime AB}~=~K^{AB}+\frac{\omega}{2}\left(\sum_{i=1}^n x_i^{AB}\right)~.
\end{equation}
Redefining one generator however changes commutation relations with others and leads to necessary modifications. An example is the modification of the dilaton which comes from the commutation relation 
\begin{equation}
        \left[P_{EF}, K^{\prime AB}\right]~=~ \frac{1}{2}\delta^{[A}_{[E}\left(M^{B]}_{F]}+\frac{1}{4}\delta^{B]}_{F]}D\right)-\frac{1}{8}\delta^{[A}_{[E}\delta^{B]}_{F]}\left(D-n\omega \right)~.
\end{equation}
The first bracket is the traceless Lorentz generator and the second leads to dilaton modification
\begin{equation}
        D^{\prime}~=~D-\omega n~.
\end{equation}
This obviously annihilates the ansatz \eqref{ansatzII}.

At this point we note that the transformation properties of our invariants are  different from those of 6D maximally supersymmetric  $\mathcal{N}=(1,1)$ SYM theory. This is because we are looking for invariants of the 6D dual superconformal algebra in analogy with the $D=4$ $\mathcal{N}=4$ SYM theory rather than objects that necessarily dimensionally reduce to the $D=4$ $\mathcal{N}=4$ SYM amplitudes. In other words we require the transformation property \eqref{amplitude_inversion} on the full amplitude $\mathcal{A}$ as opposed to  the $\mathcal{N}=(1,1)$ case \cite{Plefka14}, where the same transformation rule is  satisfied just for  the function $f_n$,
\begin{equation}
 I[f_n]~=~\left[\displaystyle\prod_{i=1}^{n}\left(x_i^2\right)\right]f_n~,
\end{equation}
and consequently the full amplitude transforms as
\begin{equation}
    I\left[\mathcal{A}_n^{\mathcal{N}=(1,1)}~\right] ~=~ (x_1^2)^2\left[\prod_{i=1}^n (x_i^2)\right] \mathcal{A}_n^{\mathcal{N}=(1,1)}~.
\end{equation}

\subsection{Example of 4-point functions} 
   
Let us now consider an explicit example of a four-point function. Apart from assumptions related to ansatz \eqref{ansatzII} we assume Lorentz invariance, which implies that the function $g_n$ is a function of $x_{ij}^2$  only. It is also reasonable to assume factorizability for a tree four-point function, i.e.\ the function $g_n(x)$ can be written in the product form
\begin{equation}
\label{ansatz_g} 
g_4(x)~=~\prod_{1\leq i<j\leq4} \left[(x_{ij})^{2}\right]^{\alpha_{ij}},
\end{equation}
with 6 real $\alpha_{ij}$-parameters.
The inversion rule is
\begin{equation}
I[g_4(x)]~=~g_4(x)\prod_{1\leq i<j\leq 4}(x_i^2x_j^2)^{-\alpha_{ij}}
~=~\frac{g_4(x)}{(x_1^2)^2}(x_1^2x_2^2x_3^2x_4^2)^{\omega}.
\end{equation}
Comparing powers, we get 4 linear equations with 6 unknowns
\begin{equation}
\label{setn4}
\begin{array} {rcl}
\alpha_{12}+\alpha_{13}+\alpha_{14}&=&2-\omega~,\\
\alpha_{12}+\alpha_{23}+\alpha_{24}&=&-\omega~,\\
\alpha_{13}+\alpha_{23}+\alpha_{34}&=&-\omega~,\\
\alpha_{14}+\alpha_{24}+\alpha_{34}&=&-\omega~.
\end{array}
\end{equation}
Next we put all nearest neighbors $\alpha_{12}$, $\alpha_{23}$, $\alpha_{34}$ and $\alpha_{14}$ to $0$ (because we are dealing with massless particles), which decreases the number of unknowns to 2. It is easy to see that  the set of equations \eqref{setn4} has no solution. Therefore we conclude that there is no four-point function of the form \eqref{ansatzII}.

\subsection{Example of 6-point functions}
 
We again assume Lorentz invariance and factorizability of six-point functions, which lead to the ansatz
\begin{equation}
\label{ansatz_g6} 
g_6(x)~=~\prod_{1\leq i<j\leq6} \left[(x_{ij})^{2}\right]^{\alpha_{ij}}~.
\end{equation}
The inversion rule \eqref{g_inversion} for $n=6$ takes the form
\begin{equation}
I[g_6(x)]~=~g_6(x)\prod_{1\leq i<j\leq 6}(x_i^2x_j^2)^{-\alpha_{ij}}
~=~\frac{g_6(x)}{(x_1^2)^2}(x_1^2x_2^2x_3^2x_4^2x_5^2x_6^2)^{\omega}.
\end{equation}
Comparing powers, we get 6 linear equations with 15 unknowns
\begin{equation}
\begin{array} {rcl}
\alpha_{12}+\alpha_{13}+\alpha_{14}+\alpha_{15}+\alpha_{16}&=&2-\omega~,\\
\alpha_{12}+\alpha_{23}+\alpha_{24}+\alpha_{25}+\alpha_{26}&=&-\omega~,\\
\alpha_{13}+\alpha_{23}+\alpha_{34}+\alpha_{35}+\alpha_{36}&=&-\omega~,\\
\alpha_{14}+\alpha_{24}+\alpha_{34}+\alpha_{45}+\alpha_{46}&=&-\omega~,\\
\alpha_{15}+\alpha_{25}+\alpha_{35}+\alpha_{45}+\alpha_{56}&=&-\omega~,\\
\alpha_{16}+\alpha_{26}+\alpha_{36}+\alpha_{46}+\alpha_{56}&=&-\omega~.
\end{array}
\end{equation}
Putting all 6 nearest neighbours to 0 there remain 9 unknowns for 6 equations. Therefore we should expect 3 free parameters. The general solution has indeed 3 free parameters $\alpha_{14}$, $\alpha_{25}$ and $\alpha_{36}$: 
\begin{equation}
\begin{array}{rcl}
\alpha_{24}&=&\frac{1}{2}\left(-\alpha_{14}-\alpha_{25}+\alpha_{36} -\omega\right)~=~\alpha_{15}-1~,\\
\alpha_{26}&=&\frac{1}{2}\left(+\alpha_{14}-\alpha_{25}-\alpha_{36} -\omega\right)~=~\alpha_{35}+1~,\\
\alpha_{46}&=&\frac{1}{2}\left(-\alpha_{14}+\alpha_{25}-\alpha_{36} -\omega\right)~=~\alpha_{13}-1~,
\end{array}
\end{equation}
The full solution parametrized by  $\alpha_{14}$, $\alpha_{25}$ and $\alpha_{36}$ can be written in the form:
\begin{equation}
\begin{array}{rcl} 
g_6\left(x_{kl}\right)
&=&\frac{x_{13}^2x_{15}^2}{x^2_{35}}
\left(x^2_{13}x^2_{15}x^2_{35}x^2_{24}x_{26}^2x^2_{46}\right)^{-\omega/2}
\left(\left(x^2_{14}\right)^{2}\frac{x^2_{35}}{x^2_{13}x^2_{15}}\frac{x^2_{26}}{x^2_{24}x^2_{46}}\right)^{\alpha_{14}/2} \cr
&&\times\left(\left(x^2_{25}\right)^{2}\frac{x^2_{13}}{x^2_{15}x^2_{35}}\frac{x^2_{46}}{x^2_{24}x^2_{26}}\right)^{\alpha_{25}/2}
\left(\left(x^2_{36}\right)^{2}\frac{x^2_{15}}{x^2_{13}x^2_{35}}\frac{x^2_{24}}{x^2_{26}x^2_{46}}\right)^{\alpha_{36}/2}~.
\label{6ptsoldiag}
\end{array}
\end{equation}
There is however another parametrization, where the dual conformal properties of the function $g_6$ are manifest. We can rewrite the expressions in parentheses with help of the dual conformal cross-ratios
\begin{equation}
u_1~:=~\frac{x_{13}^2x^2_{46}}{x^2_{14}x^2_{36}}~, \qquad u_2~:=~\frac{x_{24}^2x^2_{15}}{x^2_{14}x^2_{25}}~, \qquad
u_3~:=~\frac{x_{35}^2x^2_{26}}{x^2_{36}x^2_{25}}~,
\end{equation}
and the 6 point function takes the form
\begin{equation}
g_6\left(x_{kl}\right)~=~\frac{x_{13}^2x_{15}^2}{x^2_{35}}
\left(x^2_{14}x_{26}^2x^2_{35}\right)^{-\omega}
u_1^a u_2^b u_3^c~,\label{g6}
\end{equation}
where parameters $a$, $b$ and $c$ are linear combinations of parameters $\alpha_{14}$, $\alpha_{25}$ and $\alpha_{36}$.

We have found the solution to the dual conformal constraints however there remain unfixed parameters $a$, $b$ and $c$.  We should introduce some physics to put some limits on parameters. The first contraints come from the pole structure of amplitudes. Since we are dealing with tree-level  functions, we should consider simple poles only. This leads to the following set of inequalities between parameters   $a$, $b$, $c$ and $\omega$:
\begin{equation}
\label{nerovnice1}
a+b+c\leq -\omega~, \quad a+b\geq\omega-3~, \quad b+c\geq2\omega-1~, \quad a+c\geq2\omega-1~,
\end{equation}
\begin{equation}
\label{nerovnice2}
a\geq-1~,\quad b\geq-1~,\quad c\geq\omega~, \quad a+c\leq1~, \quad b+c\leq1~, \quad a+b\leq 1-\omega~.
\end{equation}
Combining all inequalities in \eqref{nerovnice1} we get an upper bound on $\omega$
\begin{equation}
\label{omega_bound}
\omega \leq \frac{5}{7}~.
\end{equation}
Taking into account the known amplitudes with dual conformal invariance (covariance) \cite{Drummond08}, \cite{Plefka09}, \cite{Huang_Lipstein10}, we assume that parameter $\omega$ can be integer or half-integer valued. The equation \eqref{omega_bound}  tells us that there are only two non-negative values  for $\omega~=~ \{0, 1/2\}$. The set of equations \eqref{nerovnice1} and \eqref{nerovnice2} has $5$ solutions for $\omega~=~5$ and $10$ solutions for $\omega~=~0$. These solutions are listed in Appendices \ref{omega1/2} and \ref{omega0}. We should note that there are solutions for negative $\omega$ also. The number of solutions for equations \eqref{nerovnice1} and \eqref{nerovnice2} for the first few negative values of $\omega$ is given in Table \ref{tabulka2}.
\begin{table}
\caption{Number $\#$ of solutions to inequalities \eqref{nerovnice1} and \eqref{nerovnice2}  for different  $\omega$.}
\centering
\begin{tabular}{|c||c|c|c|c|c|c|c|c|}
\hline $\omega$ &1 & $\frac{1}{2}$ & 0 & $-\frac{1}{2}$ & -1 & $-\frac{3}{2} $ &-2& ~\ldots \\ \hline
 $\#$ & 0 & 5 & 10  & 71& 26 & 179 & 48 & ~\ldots\\  \hline
\end{tabular}
\label{tabulka2} 
\end{table}

There are further constraints on scattering amplitudes  e.g.\ some relations among the scattering amplitudes. A crucial relation for color-stripped amplitudes is so-called $U\left(1\right)$ decoupling identity \cite{Elvang13}. We can construct the full six-point function for appropriate $\omega$ by taking a linear combination of individual building blocks, cf. expressions \eqref{omega1/2} for $\omega~=~1/2$ and \eqref{omega0} for $\omega~=~0$.  There is however \textit{no} linear combination of building blocs for $\omega~=~1/2$ nor $\omega~=~0$. We conclude that there are no such functions satisfying $U(1)$ decoupling identity with non-negative $\omega$.

\section{Conclusion and discussion}
We constructed a representation of the 6D dual superconformal algebra which is an analogue of the 4D dual superconformal symmetry of $\mathcal{N}=4$ SYM. The requirement of preserving constraints \eqref{constraints6D} leads to new variables $y_i^{IJ}$ and a new third constraint \eqref{C_3}.  
The generators form a Lie algebra and they commute with the three commuting constraints modulo these constraints without imposing any new constraints. 
\\
\indent The introduction of superindices $\mathcal{A}$ leads to significant simplification of the form of the generators. There is however one subtlety.  Using superized formalism one should be careful when using the generators without indices or traceless generators. An example is the superized dilatation operator $\mathds{D}$ that is not the physical dilatation generator $D$ that annihilates the scattering amplitudes, however it is a difference of the dilatation generator and hypercharge: $\mathds{D}=D-B$. The same holds for superized dilatation generator at the level of the superconformal algebra.\\
\indent An interesting check is the dimensional reduction to massless 4D theory which leads to the generators of $psu(2,2|4)$ when restricting the space of functions to scattering amplitudes. There are two exceptions. The first is the dual Lorentz generator. Comparing the known results (for example in \cite{Plefka09}) and our result we find that our result is not symmetrized in spinor indices. The solution to this problem presumably is to introduce the "full" model containing both spinors $\lambda^{Aa}_i$ and $\tilde{\lambda}_{iAa}$. However it does not seem to be straightforward to generalize in this direction. The second exception is the dual $R$-symmetry generator. Its dimensional reduction gives the non-traceless part of the 4D generator only. The dimensional reduction to massless 3D and the special form of Grassmann variable $\eta_i^{Ia}$ lead directly to the generators of the $osp(6|4)$ dual superconformal symmetry of $\mathcal{N}=6$ ABJM. \\
\indent It is possible to find an analogue of the 4D central charge in the 6D formalism. This analogue of central charge is proportional of the Casimir operator of $su(2)$ algebra and therefore is quadratic in $h_{ab}$.\\
\indent We worked out in section \eqref{examples} some examples of functions that are invariant under a subalgebra of the 6D dual superconformal algebra. We used dual conformal inversion to specify functions of our ansatz. Imposing further physical constraints on such candidates we were able to find building blocks for several non-negative values of the parameter $\omega$. However there are no linear combinations of building blocks that would satisfy relations among the scattering amplitudes, e.g.\ a $U(1)$ decoupling identity.\\
\indent Although the examples of the invariants constructed in this paper are fundamentally different from those of 6D $\mathcal{N}=(1,1)$ SYM, there remain open questions for future work, e.g.\  whether invariants of the 6D dual superconformal algebra can be written as residues of some Grassmannian integral.  Such relation between $R$-invariants as residues of Grassmanian integral and scattering amplitudes played an important role in understanding the Grassmannian formulation for scattering amplitudes  of massless  $D=4$ $\mathcal{N}=4$ SYM \cite{Mason09}. Although there is no similar 6D formula so-far, the promising direction seems to be symplectic Grassmannians introduced e.g. in \cite{Schwarz19} and \cite{cachazo18}.\\

\textbf{{Acknowledgements}}
\\
The work of K.B. is supported by the Grant Agency of the Czech Republic (GACR) under the grant P201/12/G028. We would like to thank Y.-t. Huang, A. Lipstein and P. Novosad for helpful discussions.

\appendix 
\makeatletter \@addtoreset{equation}{section} \makeatother
\def\theequation{\thesection.\arabic{equation}}
\section{Conventions} 
\label{conventions}
\indent The symmetrization and antisymmetrization are always meant without one half and are related to \textit{two} closest indices surrounded by the parentheses. To avoid confusion we stress that we only symmetrize or antisymmetrize \textit{two} indices in this paper, e.g. 
\be
o^{[AB}o^{C]D}=o^{AB}o^{CD}-o^{CB}o^{AD}, \qquad o^{\{AB}o^{C\}D}=o^{AB}o^{CD}+o^{CB}o^{AD}~,
\ee
as opposed to (anti)symmetrization of three or more indices.\\
\indent When defining derivatives, we use a shorthand for the Kronecker delta, e.g. 
\begin{equation}
\delta_{AB}^{CD} =\delta_A^C\delta_B^D, \qquad \delta_{[AB]}^{CD}=\delta_{AB}^{[CD]},\qquad  \delta_{AB}^{[CD]}=\delta_{A}^{C}\delta_{B}^{D}-\delta_{A}^{D}\delta_{B}^{C}~,
\end{equation}
in the case of 6D spinor indices.
This notation can be used also for indices of different types, for example $R$-symmetry and spinor
\be
\delta^{IA}_{JB}=\delta^{I}_J\delta^A_B~.
\ee
\indent Our epsilon symbol convention is
\begin{equation}
\epsilon^{\alpha\beta}\epsilon_{\beta\gamma}=\delta_{\gamma}^{\alpha}, \qquad \epsilon_{\alpha\beta}=\left(
\begin{array}{cc}
0&1\\
-1&0
\end{array}
\right), \qquad \epsilon^{\alpha\beta}=\left(
\begin{array}{cc}
0&-1\\
1&0
\end{array}\right)~.
\end{equation}

\subsection{3D}
The Lorentz group in 3D is $SO^+(1,2) \cong SL(2, \mathbb{R})/\mathbb{Z}_2$. The vector is symmetric in spinor indices. The null momenta can be written 
\be
p_i^{\alpha\beta}=\lambda_i^{\alpha}\lambda_i^{\beta}~.
\ee
We can raise and lower spinor indices with help of the epsilon symbol
\be
\lambda^{\alpha}_i=\epsilon^{\alpha\beta}\lambda_{i\beta}~, \qquad \lambda_{i\alpha}=\epsilon_{\alpha\beta}\lambda^{\beta}_i~.
\ee
In this paper we use the following shorthands for 3D derivatives
\be
\partial_{i\alpha}=\frac{\p}{\p \lambda_i^{\alpha}}, \quad \partial_{iI}=\frac{\partial}{\p \eta^I_i},
\ee
\begin{equation*}
\p_{i\alpha\beta}=\frac{\p}{\p x_i^{\alpha\beta}}, \quad \partial_{iI\alpha}=\frac{\p}{\p \theta_i^{I\alpha}}, \quad \p_{iIJ}=\frac{\p}{\p y_i^{IJ}}~.
\end{equation*}
The definitions of 3D derivatives are
\be
\partial_{\alpha}\lambda^{\beta}=\delta_{\alpha}^{\beta}~, \quad \partial_{I}\eta^{J}=\delta_I^J~, 
\ee
\begin{equation*}
\partial_{\alpha\beta}x^{\gamma\delta}=\frac{1}{2}\delta_ {\{\alpha\beta\}}^{\gamma\delta}~, \quad \partial_{I\alpha}\theta^{J\beta}=\delta_{I}^J\delta_{\alpha}^{\beta}~, \quad \partial_{IJ}y^{KL}=\frac{1}{2}\delta_{[IJ]}^{KL}~.
\end{equation*}

\subsection{4D}
The Lorentz group in 4D is $SO^+(1,3)$ and the corresponding complexified spin group $Spin(4,\mathbb{C})$ is isomorphic to $SL(2,\mathbb{C}) \times SL(2,\mathbb{C}) = SU(2)_{\mathbb{C}} \times SU(2)_{\mathbb{C}}$ . Therefore the vector has two spinor indices $\alpha$ and $\dot{\alpha}$. Every null momenta can then be written 
\be
p_i^{\alpha\dot{\beta}}=\lambda_i^{\alpha}\tilde{\lambda}_i^{\dot{\beta}}~.
\ee
We can raise and lower spinor indices with help of the dotted and undotted epsilon symbols
\be
\lambda^{\alpha}_i=\epsilon^{\alpha\beta}\lambda_{i\beta}~, \quad \lambda_{i\alpha}=\epsilon_{\alpha\beta}\lambda^{\beta}_i~, \qquad  \tilde{\lambda}_i^{\dot{\alpha}}=\epsilon^{\dot{\alpha}\dot{\beta}}\tilde{\lambda}_{i\dot{\beta}}~,  \quad \tilde{\lambda}_{i\dot{\alpha}}=\epsilon_{\dot{\alpha}\dot{\beta}}\tilde{\lambda}^{\dot{\beta}}_i~.
\ee
In this paper we use the following shorthands for 4D derivatives 
\be
\partial_{i\alpha}=\frac{\p}{\p \lambda_i^{\alpha}}, ~~ \partial_{i\dot{\alpha}}=\frac{\partial}{\partial \tilde{\lambda}_i^{\dot{\alpha}}}~,~~ \partial_{iI}=\frac{\partial}{\p \eta^I_i},~~ \p_{i\alpha\dot{\beta}}=\frac{\p}{\p x_i^{\alpha\dot{\beta}}}, ~~ \partial_{iI\alpha}=\frac{\p}{\p \theta_i^{I\alpha}}.
\ee
The definitions of 4D derivatives are
\be
\partial_{\alpha}\lambda^{\beta}=\delta_{\alpha}^{\beta}~, \quad \partial_{\dot{\alpha}}\tilde{\lambda}^{\dot{\beta}}=\delta_{\dot{\alpha}}^{\dot{\beta}}~,  \quad \partial_{I}\eta^{J}=\delta_I^J~, 
\ee
\begin{equation*}
\partial_{\alpha\dot{\beta}}x^{\gamma\dot{\delta}}=\delta_ {\alpha\dot{\beta}}^{\gamma\dot{\delta}}~, \quad \partial_{I\alpha}\theta^{J\beta}=\delta_{I\alpha}^{J\beta}~.
\end{equation*}

\subsection{6D}
The Lorentz group in 6D was already discussed in chapter \ref{6Dspinors}. In chapter \ref{6Dconstruction} we argued that we use only one chirality $\lambda^{Aa}_i$ of spinors.  Let us briefly recall that any null-vector can be written in terms of spinors 
\be
p^{AB}_i=\lambda^{Aa}_i\lambda^B_{ia}~.
\ee
There is no way how to raise or lower single spinor index in 6D therefore we can only raise or lower the single little group index with the epsilon symbol
\be
\lambda^{Aa}_i=\epsilon^{ab}\lambda^A_{ib}~, \qquad \lambda^A_{ia}=\epsilon_{ab}\lambda^{Ab}_i~.
\ee
Used abbreviations for 6D derivatives are
\be
\p_{iAa}=\frac{\p}{\p \lambda^{Aa}_i}, \quad \p_{iIa}=\frac{\p}{\p \eta_i^{Ia}},
\ee
\begin{equation*} 
\p_{iAB}=\frac{\p}{\p x_i^{AB}}, \quad \partial_{iIA}=\frac{\p}{\p \theta^{IA}_i}, \quad \partial_{iIJ}=\frac{\p}{\p y_i^{IJ}}~.
\end{equation*}
The definitions of 6D derivatives are
\begin{equation}
\partial_{Aa}\lambda^{Bb}=\delta_{Aa}^{Bb}~, \quad \partial_{Ia}\eta^{Jb}=\delta^{Jb}_{Ia}~, 
\end{equation}
\begin{equation*} 
\partial_{AB}x^{CD}=\frac{1}{2}\delta_{[AB]}^{CD}, \quad \partial_{IA}\theta^{JB}=\delta^{JB}_{IA}~, \quad \partial_{IJ}y^{KL}=\frac{1}{2}\delta^{KL}_{\{IJ\}}~.
\end{equation*}

\section{Dirac Derivative}
\label{appdiracderiv}

\vspace{5mm}
\noindent
In this appendix, we spell out in detail, various constrained differentiations performed in the main text, such as, e.g., differentiation wrt.\ symmetric or antisymmetric matrices.

\vspace{5mm}
\noindent
{\em Setup}. Let there be given an $m$-dimensional manifold $M$ with coordinates $(x^1, \ldots, x^m)$. Let there be given an $n$-dimensional physical submanifold $N$ with physical coordinates $(y^1, \ldots, y^n)$. Let there be given $m-n$ independent constraints
\begin{equation}
 \chi^1(x)~\approx~ 0,\quad  \ldots,\quad  \chi^{m-n}(x)~\approx~ 0, \label{diracconstr01}
\end{equation}
which defines the physical submanifold $N$. [For the $\approx$ symbol, see footnote \ref{footnote4}.] Assume that 
\begin{equation}
(y^1, \ldots, y^n, \chi^1, \ldots,\chi^{m-n})\label{diraccoord01}
\end{equation}
constitutes a coordinate system for the whole manifold $M$.

\vspace{5mm}
\noindent
{\em Dirac derivative}. In analogy with the Dirac bracket, let us introduce a {\em Dirac derivative}
 {\footnote{The subscript $D$ is a shorthand for Dirac, while subscripts $y$ and $\chi$  mean that these variables are kept fixed during the differentiation.}}
\begin{equation}
\label{diracderiv01}
\left(\frac{\partial}{\partial x^i}\right)_{\! D} 
~:=~\frac{\partial}{\partial x^i} 
-\sum_{a=1}^{m-n}\frac{\partial \chi^a}{\partial x^i}\left(\frac{\partial }{\partial \chi^a}\right)_{\! y}
~=~\sum_{\alpha=1}^n\frac{\partial y^{\alpha}}{\partial x^i} \left(\frac{\partial }{\partial y^{\alpha}}\right)_{\! \chi}
\end{equation}
\begin{equation*}
 i~\in~\{1,\ldots, m\}, 
\end{equation*}
that projects onto the physical submanifold $N$
\begin{equation}
\label{diracderivproperty01}
\left(\frac{\partial}{\partial x^i}\right)_{\! D} y^{\alpha}~=~\frac{\partial y^{\alpha}}{\partial x^i} , \qquad \left(\frac{\partial}{\partial x^i}\right)_{\! D} \chi^a~=~0, 
\end{equation}
\begin{equation*}
i~\in~\{1,\ldots, m\},\qquad \alpha~\in~\{1,\ldots, n\},\qquad a~\in~\{1,\ldots, m\!-\!n\}.
\end{equation*}

\vspace{5mm}
\noindent
{\em Does Dirac derivatives commute?} Does the commutator 
\begin{equation}
\left[ \left(\frac{\partial}{\partial x^i} \right)_{\! D}, 
\left(\frac{\partial}{\partial x^j}\right)_{\! D}\right]
~=
\end{equation}
\begin{equation*}
=~\sum_{\alpha,\beta=1}^n\frac{\partial y^{\alpha}}{\partial x^i}\left[\left(\frac{\partial }{\partial y^{\alpha}}\right)_{\! \chi},
~\frac{\partial y^{\beta}}{\partial x^j}\right]
\left(\frac{\partial }{\partial y^{\beta}}\right)_{\! \chi} - (-1)^{|i||j|}(i\leftrightarrow j)
~\stackrel{?}{\approx}~0
\label{diraccom01}
\end{equation*}
vanishes weakly?
Not necessarily. But if the coordinate transformation $x^i \leftrightarrow (y^{\alpha}, \chi^a)$ is linear, then the Dirac derivatives indeed commute. Fortunately, this is the case for our various applications in the main text. It is notably {\em not} the case for the quadratic constaints \eqref{constraints_super}, which instead are dealt with explicitly in the full space.

\vspace{5mm}
\noindent
{\em Generalizations}.  For starters, it is actually enough if \eqref{diraccoord01} is a global coordinate system in a tubular neighborhood of $N$ but not necessarily on the whole manifold $M$. Additional complications arise if the coordinates and/or constraints are not globally defined on a tubular neighborhood of $N$.

\vspace{5mm}
\noindent
{\em Reparametrizations of the constraints.} Assume that there exists a second coordinate system
\begin{equation}
(\tilde{y}^1, \ldots, \tilde{y}^n, \tilde{\chi}^1, \ldots,\tilde{\chi}^{m-n})
\label{diractildecoord01}
\end{equation}
(which we adorn with tildes), such that
\begin{equation}
\tilde{y}^{\alpha}~=~f^{\alpha}(y), \qquad  
\tilde{\chi}^a ~=~ g^a(y,\chi)~\approx~0.\label{diractildecoord02}
\end{equation}
This implies that 
\begin{equation}
\left(\frac{\partial }{\partial \chi^a}\right)_{\! y}
~=~ \left(\frac{\partial \tilde{\chi}^b}{\partial \chi^a}\right)_{\! y}  \left(\frac{\partial }{\partial \tilde{\chi}^b}\right)_{\! \tilde{y}}, \qquad
\left(\frac{\partial }{\partial y^{\alpha}}\right)_{\! \chi}
~\approx~\left(\frac{\partial \tilde{y}^{\beta}}{\partial y^{\alpha}}\right)_{\! \chi} \left(\frac{\partial }{\partial \tilde{y}^{\beta}}\right)_{\! \tilde{\chi}},
\label{diractildederiv01}
\end{equation}
i.e.
\begin{equation}
\Delta_{\chi}~:=~{\rm span}\left\{\left(\frac{\partial }{\partial \chi^1}\right)_{\! y}, \ldots ,\left(\frac{\partial }{\partial \chi^{n-m}}\right)_{\! y}\right\} ~\subseteq~ TM
\label{diracdistrib01}
\end{equation} 
is an involutive distribution, while 
\begin{equation}
\Delta_y~:=~{\rm span}\left\{\left(\frac{\partial }{\partial y^1}\right)_{\! \chi}, \ldots ,\left(\frac{\partial }{\partial y^n}\right)_{\! \chi}\right\} ~\subseteq~ TM
\label{diracdistrib02}
\end{equation} 
is a weak distribution.

\vspace{5mm}
\noindent
One may show that the Dirac derivative and its commutators
\begin{equation} 
\label{diractildederiv02}
\left(\frac{\partial}{\partial x^i}\right)^{\sim}_{\! D}~\approx~\left(\frac{\partial}{\partial x^i}\right)_{\! D}, 
\end{equation}
\begin{equation*} 
\left[ \left(\frac{\partial}{\partial x^i} \right)^{\sim}_{\! D}, 
\left(\frac{\partial}{\partial x^j}\right)^{\sim}_{\! D}\right]~\approx~
\left[ \left(\frac{\partial}{\partial x^i} \right)_{\! D}, 
\left(\frac{\partial}{\partial x^j}\right)_{\! D}\right],
\end{equation*}
[wrt. the tilde and the untilde coordinate systems \eqref{diractildecoord01} and \eqref{diraccoord01}, respectively] agree weakly. This shows that the Dirac derivative \eqref{diracderiv01} is a geometric construction.

\vspace{5mm}
\noindent
{\em Subsubmanifold}. Given a $p$-dimensional physical subsubmanifold $P$ with physical coordinates $(z^1,\ldots,z^p)$. Let there be given $n-p$ independent constraints
\begin{equation}
\phi^1(y)~\approx~ 0,\quad  \ldots,\quad  \phi^{n-p}(y)~\approx~ 0, \label{diracconstr02}
\end{equation}
which defines the physical submanifold $P$. Assume that 
\begin{equation}
(z^1, \ldots, z^p, \phi^1, \ldots,\phi^{n-p}) \label{diraccoord02}
\end{equation}
constitutes a coordinate system for the submanifold $N$. One may show that 
\begin{equation}
\label{diracdiracderiv01} 
\left(\frac{\partial}{\partial x^i}\right)^{\!(P)}_{\! D} 
~=~\left(\frac{\partial}{\partial x^i}\right)^{\!(N)}_{\! D} 
- \sum_{a=1}^{n-p}\left(\frac{\partial \phi^a}{\partial x^i}\right)^{\!(N)}_{\! D}\left(\frac{\partial }{\partial \phi^a}\right)_{\!z}, 
\end{equation}
\begin{equation*}
 i~\in~\{1,\ldots, m\}. 
\end{equation*}

This shows that the Dirac derivative construction behaves naturally wrt.\ further constraints. 

\vspace{5mm}
\noindent
{\em Examples}. A typical example of a subsubmanifold Dirac construction is when certain blocks of an (anti)symmetric matrix (which by itself can be viewed as a Dirac construction) are additionally put to zero. E.g.\ this is the situation of $y_i^{IJ}$ variables in chapter \ref{DimRed}. These variables are symmetric in indices $I$ and $J$ which can be viewed as a constraint that projects arbitrary $y^{IJ}_i$ onto the submanifold of symmetric matrices. Further constraints arise in restricting to the $\mathcal{N}=4$ subspace where all $y_i^{IJ}\approx y_{i+1}^{IJ}$ are identified wrt.\ particle index $i$, see section \ref{DimReduction4D}. This defines in this example the subsubmanifold.

\section{Examples of 6-point functions}
In this appendix we list examples of invariants of 6D dual superconformal algebra based on the ansatz \eqref{ansatzII}. 
 \subsection{Example of 6-point function with $\omega=\frac{1}{2}$}

Only 5 terms have no more than simple pole singularities:

\begin{eqnarray}
\label{omega1/2}
(a,b,c) ~=~(-1/2,-1,1)&:& \frac{x^2_{14}}{(x^2_{36})^{1/2}}  
  \frac{(x^2_{26})^{1/2}}{(x^2_{46})^{1/2}x^2_{24}} \frac{(x^2_{13})^{1/2}}{(x^2_{35})^{1/2}} ~,\cr\cr
(a,b,c) ~=~(-1/2,-1/2,1/2)&:& \frac{(x^2_{14})^{1/2}}{(x^2_{24})^{1/2}(x^2_{46})^{1/2}}
  \frac{(x^2_{13})^{1/2}(x^2_{15})^{1/2}}{x^2_{35}}  ~,\cr\cr
(a,b,c) ~=~(-1,-1,1)&:&   \frac{(x^2_{14})^{3/2}}{(x^2_{35})^{1/2}}  
  \frac{(x^2_{26})^{1/2}}{x^2_{24}x^2_{46}}~,\cr\cr
(a,b,c) ~=~(-1,-1/2,1)&:& \frac{x^2_{14}}{(x^2_{25})^{1/2}}  
  \frac{(x^2_{26})^{1/2}}{(x^2_{24})^{1/2}x^2_{46}} \frac{(x^2_{15})^{1/2}}{(x^2_{35})^{1/2}} ~,\cr\cr
(a,b,c) ~=~ (-1,-1,3/2)&:&  
 \frac{(x^2_{14})^{3/2}}{(x^2_{25})^{1/2}(x^2_{36})^{1/2}}  
  \frac{x^2_{26}}{x^2_{24}x^2_{46}}~,
\end{eqnarray}

\subsection{Example of 6-point function with $\omega=0$}

Only 10 terms have no more than simple pole singularities:
\begin{eqnarray}
\label{omega0}
(a,b,c) ~=~(-1, -1, 0)&:& \frac{(x^2_{14})^{2}x^2_{25}x^2_{36}}{x^2_{24}x^2_{46}x^2_{35}}  ~,\quad \cr\cr
(a,b,c) ~=~(-1,-1,1)&:&\frac{ (x^2_{14})^2x^2_{26}}{x^2_{24}x^2_{46}}  ~,\cr\cr
(a,b,c) ~=~(-1,-1,2)&:& \frac{(x^2_{14})^2}{x^2_{25}x^2_{36}} \frac{(x^2_{26})^2 x^2_{35}}{x^2_{24}x^2_{46}} ~,\cr\cr
(a,b,c) ~=~(-1,0,0)&:&\frac{ x^2_{14}x^2_{36}x^2_{15}}{x^2_{46}x^2_{35}}  ~,\cr\cr
(a,b,c) ~=~(-1,0,1)&:& \frac{x^2_{14}}{x^2_{25}} \frac{x^2_{26}x^2_{15}}{x^2_{46}}  ~,\cr\cr
(a,b,c) ~=~(-1,1,0)&:& \frac{x^2_{36}}{x^2_{25}}\frac{x^2_{24}}{x^2_{46}}\frac{(x^2_{15})^{2}}{x^2_{35}}   ~,\cr\cr
(a,b,c)  ~=~(0,-1,0)&:&\frac{ x^2_{14}x^2_{25}x^2_{13}}{x^2_{24}x^2_{35}}  ~,\cr\cr
(a,b,c) ~=~(0,-1,1)&:& \frac{x^2_{14}}{x^2_{36}} \frac{x_{26}^2x_{13}^2}{x^2_{24}}  ~,\cr\cr
(a,b,c) ~=~(0,0,0)&:& \frac{x^2_{13}x^2_{15}}{x^2_{35}}  ~,\cr\cr
(a,b,c) ~=~(1,-1,0)&:& \frac{x^2_{25}}{x^2_{36}}\frac{x^2_{46}}{x^2_{24}}\frac{(x^2_{13})^{2}}{x^2_{35}}  ~,
\end{eqnarray}

\end{document}